\documentclass[pdftex,twocolumn,epjc3]{svjour3}          
 
\RequirePackage[T1]{fontenc}
\smartqed  
\usepackage{datetime}
\usepackage[normalem]{ulem}
\usepackage{mathtools, cuted}
\usepackage{dblfloatfix,float}
\usepackage{booktabs}
\usepackage{amsmath,epsfig,amssymb,multirow}
\usepackage{enumerate,cancel}
\RequirePackage{graphicx}
\RequirePackage{mathptmx}      
\RequirePackage{flushend}
\usepackage{cite}
\RequirePackage[numbers,sort&compress]{natbib}
\RequirePackage[colorlinks,citecolor=blue,urlcolor=blue,linkcolor=blue]{hyperref}
\let\oldhref\href
\renewcommand{\href}[2]{\oldhref{#1}{\hbox{#2}}}

\usepackage[symbol]{footmisc}

\usepackage[a4paper, inner=1.5cm, outer=1.5cm, top=2.0cm,bottom=2.5 cm, binding offset=0 cm]{geometry}
\journalname{HRI-RECAPP-2022-010}
\allowdisplaybreaks
\def\p{\partial}

\usepackage{color}
\definecolor{myred}{rgb}{0.6,0,0} 
\definecolor{myblue}{rgb}{0,0.2,0.4}
\definecolor{mygreen}{rgb}{0,0.9,0.1}
\definecolor{hc}{rgb}{.9,0.1,0.7}
\definecolor{hcout}{rgb}{.9,0.7,0.9}
\definecolor{Orange}{rgb}{1.,0.65,0.}

\allowdisplaybreaks
\def\p{\partial}

\def\l{\left}

\def\wt{\widetilde}
\def\({\left(}
\def\){\right)}
\def\[{\left[}
\def\]{\right]}

\newcommand{\ba}{\begin{array}}
	\newcommand{\ea}{\end{array}}
\newcommand{\bd}{\begin{displaymath}}
	\newcommand{\ed}{\end{displaymath}}
\newcommand{\be}{\begin{equation}}
	\makeatletter
	\newcommand*{\rom}[1]{\expandafter\@slowromancap\romannumeral #1@}
	\makeatother
	\newcommand{\ee}{\end{equation}}
\def\bt{\begin{table}}
	\def\et{\end{table}}
\def\bc{\begin{center}}
	\def\ec{\end{center}}
\def\bi{\begin{itemize}}
	\def\ei{\end{itemize}}
\def\bw{\begin{widetext}}
	\def\ew{\end{widetext}}

\def\bea{\begin{eqnarray}}
	\def\eea{\end{eqnarray}}
\def\beas{\begin{eqnarray*}}
	\def\eeas{\end{eqnarray*}}

\def\N0{\widetilde{\chi}^0}

\def\a{\alpha}

\def\d{\delta}

\def\m{\mu}

\def\ov {\overline}

\def\l{\lambda}

\begin{document}
	
	\title{ Fatjet signatures of heavy neutrinos and heavy leptons in  a left-right model with universal seesaw at the HL-LHC}
	
	\author{Atri~Dey,\thanksref{e1,ad1}~ Rafiqul Rahaman,\thanksref{e2,ad1}~ and Santosh~Kumar~Rai\thanksref{e3,ad1}}
	\thankstext{e1}{email:atridey@hri.res.in}
	\thankstext{e2}{email:rafiqulrahaman@hri.res.in}
	\thankstext{e3}{email:skrai@hri.res.in}
	\institute{\em Regional Centre for Accelerator-based Particle Physics, Harish-Chandra Research Institute,  A CI of Homi Bhabha National Institute,
		Chhatnag Road, Jhunsi, Prayagraj 211019, India\label{ad1}}
	\date{}
	\maketitle
	\begin{abstract}
		We perform a collider search for fatjet signals originating from boosted heavy neutral and charged leptons with masses between a few hundred GeV to a TeV. These heavy leptons originate from the decay of heavy gauge bosons with masses above 4 TeV in a left-right symmetric extension of the Standard Model (SM), which considers a  universal seesaw mechanism for the generation of all the SM fermion masses. The fatjet signals arise naturally in this model due to the presence of heavy seesaw partners of the SM fermions which decay to SM gauge bosons carrying large boosts. We employ substructure based variables lepton sub-jet fraction ($LSF$) and lepton mass drop ($LMD$) together with kinematic variables of fatjets to look for fatjet signals associated with non-isolated leptons. These variables help in reducing the SM backgrounds while retaining enough statistics for signal events, which leads to a robust discovery potential at the high-luminosity Large Hadron Collider (HL-LHC).
	\end{abstract}

\section{Introduction}\label{sec:intro}
Non zero neutrino mass observed in experiments~\cite{SNO:2002tuh,Super-Kamiokande:1998kpq,KamLAND:2002uet,Esteban:2020cvm,NuFit50} is one of the key hints to look for physics beyond the Standard Model (SM). Some other aspects unexplained in the SM are dark matter, hierarchy of masses in the three known fermion generations, gauge hierarchy problem, Baryogenesis, $CP$-violation, etc.  Left-right symmetric extension of the SM  (LRSM)~\cite{Mohapatra:1974gc,Mohapatra:1974hk, Senjanovic:1975rk,Senjanovic:1978ev} is one of the well motivated models of new physics, which offers an explanation for some of the experimental as well as theoretical limitations of the SM mentioned above. Left-right symmetric models resolve the issue of maximal parity violation in the weak sector and naturally accommodate right-handed neutrinos in the framework which leads to the popular seesaw mechanism for neutrino mass generation~\cite{ Minkowski:1977sc,Sawada:1979dis,Glashow:1979nm,Mohapatra:1979ia}. 
 The seesaw framework gives rise to Majorana masses to the neutrinos and the heavy neutrinos have lepton-number violating interactions. These heavy states can decay via their CP-violating Yukawa interactions to generate a lepton asymmetry. This lepton asymmetry is then partially converted to a baryon asymmetry through the SM sphaleron processes that can explain the matter-antimatter asymmetry~\cite{Kuzmin:1980yp,Kuzmin:1985mm,Frere:1992bd,Sarkar:2007er}. The LRSM models also account for $CP$-violation and resolve the strong $CP$ problem~\cite{Mohapatra:1978fy,Babu:1988mw,Babu:1989rb,Barr:1991qx,Barenboim:1996rn,Kuchimanchi:2010xs}.
 
 The  minimal left-right models include scalar triplets and a scalar bi-doublet along with right-handed SM fermions arranged in the $SU(2)_R$ as doublets~\cite{Deshpande:1990ip}. This leads to the SM neutrinos getting their eventual mass via the seesaw mechanism, while the rest of the SM fermions get mass through their Yukawa interactions with the bidoublet scalar in the usual manner. We adopt an LRSM framework that suggests that all SM fermions, including the neutrinos, get their mass from a seesaw mechanism similar to that of the neutrino~\cite{Patra:2017gak}. This can be achieved by modifying the scalar sector of LRSM with four $SU(2)$ doublet scalars, one each for the lepton and quark doublets in the left sector as well as in the right sector~\cite{Babu:2014vba,Patra:2017gak}. A big advantage of generating the fermion masses in this manner is to prevent a highly hierarchical Yukawa structure, like in the SM. The fermion sector, however, also gets modified and heavy singlet charged fermions along with heavy singlet Majorana neutrinos need to be included to achieve a  universal seesaw mechanism for the generation of all the fermion masses. In such models, even the strong CP problem is resolved without an axion if a discrete parity symmetry is imposed~\cite{Gu:2017mkm}. 
	
	This framework offers rich and interesting phenomenology whose signals can be observed at current and future colliders. 
	An interesting signal for the model will be the observation of fatjet signatures of heavy neutrinos ($\nu_i$) and exotic heavy 
	charged leptons ($E_i$) with masses around a few hundred GeV to TeV. This signal originates from the decay of the heavy right-handed 
	charged gauge boson ($W_R$) with mass above a few TeV. We note that fatjet signatures of heavy neutrinos 
 have been studied in the literature for  LRSM models  with minimal scalar sector that generate seesaw masses for neutrinos~\cite{Mitra:2016kov,Das:2017gke,ThomasArun:2021rwf} 
	at the Large Hadron Collider (LHC). However, in the universal seesaw model, the presence of 
	heavy charged fermions along with the heavy Majorana neutrinos\footnote {We shall refer to these heavy Majorana neutrinos as ``heavy neutrinos" in the remainder of the text.} lead to an even more interesting 
	picture of final states with multiple fatjets and different lepton charge multiplicities. The 
 signal originates from the production of the $W_R$ via $pp$ collision, which then decays to a heavy 
	neutrino  and a heavy charged lepton. The heavy lepton further decays to a heavy neutrino and a pair of jets via an  
	off-shell $W_R$. Each of the  heavy neutrinos (dominantly right-handed)  further decay to a lepton and a pair of jets through 
	off-shell $W_R$. The heavy neutrinos being much lighter than the $W_R$ are produced with a substantial boost, and their decay 
	products (pair of jets including a charged lepton) form boosted fatjets.  
	The emerging signal becomes a topology of three fatjets where two of them include a lepton and one is characterized by 
	two sub-jets.  This three-fatjet signal results from the following sub-process in the model,
	\begin{eqnarray}\label{eq:3-fatjet-signal}
		pp\to W_R\to E_i \nu_j,~ E_i\to \nu_k W_R^\star,\nonumber\\ ~\nu_{k}\to j j(W_R^\star) l_i^\pm,~W_R^\star\to jj.
	\end{eqnarray}  
	This LRSM framework also offers two-fatjet signals including a lepton, originating from the decay of the heavy right-handed neutral gauge boson ($Z_R$): 
	\begin{equation}\label{eq:2-fatjet-signal}
		pp\to Z_R\to \nu_i \nu_j,~\nu_{i}\to l_k^\pm jj.
	\end{equation}
	Interestingly, a possibility of a more exotic signal including  four fatjets also exists because of 
	the production of the heavy charged leptons $E_i$ produced via
	\begin{equation}\label{eq:4-fatjet-signal}
		pp\to Z_R\to E_i E_j.
	\end{equation}
	A complete account of the possible signals is given in section~\ref{sec:collider-analysis}. We note that the 
	presence of leptons in the fatjet can be a crucial identification since the leptons and jets produced from the decay of 
	the boosted heavy neutrinos are mostly non-isolated, following the standard 
	isolation criteria used by the CMS and ATLAS~\cite{CMS:1994hea,ATLAS:1994vge} Collaborations. We, therefore, use substructure-based 
	variables employed by the experimental collaborations called  lepton sub-jet fraction ($LSF$) and lepton mass drop ($LMD$) 
	to distinguish our fatjet signal from the non-reducible quantum chromodynamics (QCD) background~\cite{Brust:2014gia}. These variables ($LSF$, $LMD$) 
	will be defined later in section~\ref{sec:collider-analysis}. 
	
	The rest of the paper is organized as follows. We give a brief description of the LRSM model with a universal seesaw 
	mechanism in section~\ref{sec:model}. In section~\ref{sec:benchmarks}, we discuss the constraints on the model parameters 
	and phenomenologically allowed benchmark points for our analysis.  In section~\ref{sec:collider-analysis}, 
	we present our collider analysis and results using the variables $LSF$ and $LMD$. We conclude with our findings 
	in section~\ref{sec:conclusion}.

 \begin{table*}[h!]
		\centering
		\caption{\label{tab:qnumbers} Particle spectrum along with their quantum numbers.}
		\renewcommand{\arraystretch}{1.2}
		\begin{tabular*}{\textwidth}{@{\extracolsep{\fill}}llllll@{}}\hline
			Field & $SU(3)_C$ & $SU(2)_L$ & $SU(2)_R$ & $U(1)_{B-L}$ & ${Z}_2$ \\ \hline
			$Q_{L(R)} = \left (\begin{array}{c} u\\ d \end{array} \right )_{L(R)}$ & 3 & 2~(1) & 1~(2) & $\frac{1}{3}$ & + \\ \hline
			$l_{L(R)} = \left (\begin{array}{c} \nu \\ e \end{array} \right )_{L(R)}$ & 1 & 2~(1) & 1~(2) & -1 & + \\  \hline
			$U_L$, $U_R$ & 3 & 1 & 1 & $\frac{4}{3}$ & + \\  \hline
			$D_L$, $D_R$ & 3 & 1 & 1 & $-\frac{2}{3}$ & + \\  \hline
			$E_L$, $E_R$ & 1 & 1 & 1 & -2 & - \\ \hline
			$N_L$, $N_R$ & 1 & 1 & 1 & 0 & - \\  \hline
			$H_{L(R)Q}= \left (\begin{array}{c} H^+ \\H^0 \end{array} \right )_{L(R)Q}$ & 1 & 2~(1) & 1~(2) & 1 & + \\  \hline
			$H_{L(R)l}= \left (\begin{array}{c} H^+ \\H^0 \end{array} \right )_{L(R)l}$ & 1 & 2~(1) & 1~(2) & 1 & - \\ \hline
		\end{tabular*}
	\end{table*}	
	\section{The Model}\label{sec:model}
	A brief description of the model~\cite{Patra:2017gak} is given below.	The left-right symmetric model is based on the gauge group $SU(3)_C \times SU(2)_L \times SU(2)_R \times U(1)_{B-L}$ along 
	with an extra $Z_2$ symmetry to incorporate a lepton-specific scenario.
	The charge of a particle is defined as
	\begin{equation}
		{\cal Q}=I_{3L}+I_{3R}+\frac{B-L}{2}.
	\end{equation}
The particle spectrum of the model along with their quantum number and $Z_2$ charges are shown in Table~\ref{tab:qnumbers}.
	The  matter structure consists of three families of $SU(2)_L$ and $SU(2)_R$ quark and lepton doublets,
	\begin{eqnarray}
		Q_L&=&\left (\begin{array}{c}
			u\\ d \end{array} \right )_L \sim \left (3,2, 1, \frac13 \right ),~
		Q_R=\left (\begin{array}{c}
			u\\d \end{array} \right )_R \sim \left ( 3,1, 2, \frac13
		\right ),\nonumber \\
		l_L&=&\left (\begin{array}{c}
			\nu\\ e\end{array}\right )_L \sim\left ( 1,2, 1, -1 \right ),~
		l_R=\left (\begin{array}{c}
			\nu \\ e \end{array}\right )_R \sim \left ( 1,1, 2, -1 \right ),\nonumber\\
	\end{eqnarray}
	where the numbers in the parentheses denote the quantum numbers under
	$SU(3)_C \times SU(2)_L \times SU(2)_R \times U(1)_{B-L}$ gauge groups,  respectively. 
	For the generation of SM quark and lepton masses through a universal seesaw, the model includes 
	heavy singlet quarks $$U_L(3,1,1,\frac{4}{3}),~ U_R(3,1,1,\frac{4}{3}),~  D_L(3,1,1,-\frac{2}{3}),~ D_R(3,1,1,-\frac{2}{3});$$ 
	singlet charged leptons 
	$$E_L(1,1,1,-2),~E_R(1,1,1,-2);$$ and singlet heavy neutrinos  
	$$N_L(1,1,1,0),~N_R(1,1,1,0).$$

	The Higgs sector consists of four $SU(2)$ doublets: 
	\begin{eqnarray}
		H_{RQ}(1,1,2,1)&=&\left (\begin{array}{c}
			H_{RQ}^+ \\H_{RQ}^0 \end{array} \right ),~~
		H_{LQ}(1,2,1,1)=\left (\begin{array}{c}
			H_{LQ}^+ \\H_{LQ}^0 \end{array} \right ),~~ \notag \\
		H_{Rl}(1,1,2,1)&=&\left (\begin{array}{c}
			H_{Rl}^+ \\H_{Rl}^0 \end{array} \right ),~~
		H_{Ll}(1,2,1,1)=\left (\begin{array}{c}
			H_{Ll}^+ \\H_{Ll}^0 \end{array} \right ).~~~~~
	\end{eqnarray}
 
  Under the  $Z_2$ symmetry, the singlet leptons $E_L,~E_R,~ N_L,~ N_R$, and the lepton specific scalar doublets $H_{Ll}$ and  $H_{Rl}$ are odd, while the rest are all even.
	The $Z_2$ charge assignments allow the $H_{LQ}$ and $H_{RQ}$ to interact specifically with quarks, while  $H_{Ll}$ and $H_{Rl}$ 
	only have leptonic interactions. The neutral components of the two $SU(2)_{R/L}$ Higgs doublets acquire non-zero 
	vacuum expectation values (VEV) to break the LRSM symmetry down to $U(1)_{\rm QED}$. The VEVs are given as
	\begin{eqnarray}
		\left<H_{RQ}^0\right>=v_{RQ}/\sqrt{2},~~\left<H_{Rl}^0\right>=v_{Rl}/\sqrt{2},\nonumber\\
		~~\left<H_{LQ}^0\right>=v_{LQ}/\sqrt{2},~~\left<H_{Ll}^0\right>=v_{Ll}/\sqrt{2}
	\end{eqnarray}
	with the constraint  $v_{LQ}^2+v_{Ll}^2 = v_{EW}^2$, where  $v_{EW}=246$ GeV
	is the electroweak (EW) vacuum expectation value. The hierarchy of the VEVs  is arranged as
	\begin{equation}
		v_{RQ},v_{Rl}>>v_{LQ}>v_{Ll}.
	\end{equation}
	The VEVs of the $SU(2)_R$ scalar doublets, $v_{RQ}$ and $v_{Rl}$, are responsible for generating the mass of 
	$W_R$ and $Z_R$ gauge bosons, while the VEVs to the $SU(2)_L$ scalar doublets,$v_{LQ}$ and $v_{Ll}$, 
	break the EW symmetry and generate the SM gauge boson ($W$ and $Z$) masses.

The full Lagrangian of this model can be written as
 \begin{equation}
		\mathcal{L} = \mathcal{L}_{kinetic} + \mathcal{L}_Y + V(H), 
\end{equation}
 where $\mathcal{L}_{kinetic}$ contains the kinetic terms for the gauge boson, scalars and fermions in the model, while the Yukawa Lagrangian, $\mathcal{L}_Y$ and the scalar potential, $V(H)$ are given later in  Eq.~(\ref{eq:Yukawa}) and Eq.~(\ref{eq:scalpot}), respectively.
 
	The covariant derivatives appearing in $\mathcal{L}_{kinetic}$ can be written as,
	\begin{eqnarray}\label{eq:covd}
		{\cal D}_\mu Q_{L/R} &=& \left[ \p_\mu-i\frac{g_{L/R}}{2} \tau.W_{L/R{\mu}}-i \frac{g_V}{6} V_{\mu} \right] Q_{L/R},\nonumber\\
		{\cal D}_{\mu}l_{L/R}&=& \left[\p_{\mu}-i\frac{g_{L/R}}{2} \tau.W_{L/R{\mu}}+i\frac{g_V}{2}V_{\mu}\right]l_{L/R}, \nonumber \\
		{\cal D}_{\mu}H_{L/R} &=& \left[ \p_{\mu}-i\frac{g_{L/R}}{2} \tau.W_{L/R{\mu}}-i \frac{g_V}{2} V_{\mu}\right] H_{L/R},
	\end{eqnarray}
	where $V_{\mu}$ is the $U(1)_{B-L}$ gauge boson and $g_V$ its gauge coupling, while $W_L$, $W_R$ and $g_L$, $g_R$ are the gauge bosons and gauge couplings corresponding to the 
	$SU(2)_L$ and $SU(2)_R$ gauge groups, respectively.
	The mass of the charged gauge bosons $W_R$ and $W_L$ are given by
	\begin{equation}
		M^2_{W_R^\pm} = \frac{1}{4} g_R^2 (v_{RQ}^2 +v_{Rl}^2),~~~~ 
		M^2_{W^\pm} = \frac{1}{4} g_L^2 (v_{LQ}^2 +v_{Ll}^2).
		\label{eq:WR}
	\end{equation}
	The masses of the two neutral massive gauge bosons are given by
	\begin{eqnarray}
		M^2_{Z_R} &\simeq& \frac{1}{4} \left[(g_R^2+g_V^2) (v_{RQ}^2 +v_{Rl}^2)+ \frac{g_V^4(v_{LQ}^2+v_{Ll}^2)}{g_R^2+g_V^2} \right],\nonumber\\
		M^2_{Z} &\simeq& \frac{1}{4} (g_L^2 + g_Y^2)(v_{LQ}^2 +v_{Ll}^2)
		\label{eq:ZR}
	\end{eqnarray} 
	in the limit $v_{EW}<<v_{RQ},v_{Rl}$.  Here, $g_Y$ is  the effective SM $U(1)_Y$ gauge coupling given as
	\begin{equation}
		g_Y=\frac{g_L g_V}{\sqrt{g_L^2+g_V^2}}.
	\end{equation}
	In this model, the $Z_R$ is always heavier than $W_R$, which implies that a  strong limit on $W_R$ mass indirectly puts a stronger bound on $Z_R$ mass. We discuss the generation of masses for the fermions through the seesaw mechanism in the 
	the following subsection.
	
	\subsection{Seesaw and fermion masses}
	The quarks and leptons obtain their masses through a universal seesaw with the help of heavy singlet fermionic states. 
	The Lagrangian in this model contains the following gauge invariant Yukawa terms:
	\begin{eqnarray}
		\mathcal{L}_Y&=& \left( Y_{uL}\overline{Q}_{L} \wt{H}_{LQ} U_R + Y_{uR}\overline{Q}_{R} \wt{H}_{RQ} U_L + Y_{dL}\overline{Q}_{L} H_{LQ} D_R \right. \notag \\
		&+& Y_{dR}\overline{Q}_{R} H_{RQ} D_L 
		+ Y_{\nu L} \overline{l}_{L} \wt{H}_{Ll} N_R + Y_{\nu R}\overline{l}_{R} \wt{H}_{Rl} N_L \notag \\
		&+& Y_{eL}\overline{l}_{L} H_{Ll} E_R + Y_{eR}\overline{l}_{R} H_{Rl} E_L + M_U \ov U_L U_R + M_D \ov D_L D_R \notag \\
		&+& \left. M_E \ov E_L E_R + M_{ N} \ov N_L N_R + H.C. \right) \notag \\
		&+& M_{ L} N_L^T N_L^C + M_{ R} N_R^T N_R^C.
		\label{eq:Yukawa}
	\end{eqnarray}
	Here,  $Y_{iA}$'s are the Yukawa coupling matrices, and $M_X$'s are the heavy singlet mass terms. The $\widetilde H_{L/R}$ is the conjugate scalar defined as 
	\begin{equation}
		\widetilde H_{L/R} = i \tau_2 H^\ast _{L/R}.
	\end{equation}
	All the charged fermions, i.e., quarks and charged leptons, acquire their masses by diagonalizing $6\times 6$ matrices through a universal seesaw mechanism. Three light eigenstates among the six eigenstates are identified as the three SM states.

	The up quarks acquire their masses from the following  mass matrix
	\begin{equation}
		M_u = \begin{pmatrix}
			0&Y_{uR} v_{RQ}/\sqrt{2} \\ Y_{uL}^T v_{LQ}/\sqrt{2} & M_U
		\end{pmatrix}
	\end{equation}
	in the $\left(u,U\right)$ basis. This matrix is diagonalized by the following bi-unitary transformation:
	\begin{equation}
		M_u^{diag} = U_{uL} M_u U_{uR}^{\dagger},
		\label{eq:upmass}
	\end{equation}
	where $U_{uL}$ and $U_{uR}$  are $6\times6$ unitary matrices transforming the left-handed and right-handed fermions from the gauge basis to their mass basis. 
 The down quarks also acquire their masses through the seesaw mechanism, where the corresponding matrices are $U_{dL}$ and $U_{dR}$. For simplicity, we choose the $Y_{uL/R}$ and $Y_{dR}$ to be diagonal and keep only $Y_{dL}$ to 
	be off-diagonal so as to generate the correct\\  Cabibbo–Kobayashi–Maskawa (CKM) mixing matrix for the SM quarks.
	The $6\times 6$ mixing matrix 
	\begin{equation}
		U_{L}^{CKM} = U_{uL} U_{dL}^\dagger
	\end{equation}
	contains the  SM CKM matrix in the {\em top-left} ({\em bottom-right}) $3\times3$ block if mass eigenvalues are arranged in ascending (descending) order. The mixing between the SM light quarks and the heavy quarks is negligible
	except for the top quark as it is heavy.

	The charged leptons acquire their masses from the following $6\times 6$ mass matrix:
	\begin{equation}
		M_e = \begin{pmatrix}
			0&Y_{eR} v_{Rl}/\sqrt{2} \\ Y_{eL}^T v_{Ll}/\sqrt{2} & M_E
		\end{pmatrix}.
	\end{equation} 
	The Yukawa coupling matrices ($Y_{eL/R}$) and the heavy lepton mass matrix ($M_E$) are chosen diagonal to prevent any charged lepton flavor violating interactions. The hierarchical structure of the block matrices that give the correct masses to the SM charged leptons ($e,~\mu,~\tau$) allow negligible mixing between the SM charged leptons and heavy charged leptons.  
	
	The neutrinos obtain their masses from the following $12\times 12$ mass matrix:
	\begin{equation}
		\begin{pmatrix} 
			0&Y_{\nu L}v_{Ll}/\sqrt{2}&0&0 \\Y_{\nu L}^T v_{Ll}/\sqrt{2}&M_R&0&M_N^T \\ 0&0&0&Y_{\nu R}^T v_{Rl}/\sqrt{2}\\0&M_N&Y_{\nu R} v_{Rl}/\sqrt{2}&M_L
		\end{pmatrix}
	\end{equation}
	formed in the basis $(\nu_L^\ast,N_R,\nu_R,N_L^\ast)$.
  Similar to the quark sector, all the Yukawa couplings and mass matrices are taken diagonal except $Y_{\nu L}$ that generates the Pontecorvo-Maka-Nakagawa-Sakata (PMNS) mixing matrix~\cite{Esteban:2020cvm}.

	\subsection{Scalar masses}
	The scalars obtain their masses by minimizing the following  gauge-invariant potential
	\begin{eqnarray}
		V(H)&=& \sum_{i=1}^4 \mu_{ii} H_i^\dagger H_i + \sum_{\substack{{i,j=1} \notag \\ {i \leq j}}}^4 \l_{ij} H_i^\dagger H_i H_j^\dagger H_j \notag\\
		&+& \left[\a_1 H_{LQ}^\dagger H_{Ll} H_{RQ}^\dagger H_{Rl}+\a_2 H_{LQ}^\dagger H_{Ll} H_{Rl}^\dagger H_{RQ} \right. \notag \\
		&+& \left. \mu_{12}^2 H_{LQ}^\dagger H_{Ll} + \mu_{34}^2 H_{RQ}^\dagger H_{Rl}+ H.C. \right],
	\label{eq:scalpot}
    \end{eqnarray}
	where
	\begin{equation}
		H_1 = H_{LQ},~~ H_2 = H_{Ll},~~H_3 = H_{RQ},~~H_4 = H_{Rl}.
	\end{equation}
	The terms $\mu_{12}$ and $\mu_{34}$  break the discrete $Z_2$ symmetry softly and prevent  the formation of domain walls which could otherwise 
	destabilize the model~\cite{Zeldovich:1974uw,Kibble:1976sj}. 
 
	The Higgs boson spectrum consists of four $CP$-even states, 
	two $CP$-odd states, and two charged Higgs bosons. 
 A major distinctive feature from the more widely studied minimal LRSM model which includes scalar triplets, is the absence of a doubly charged Higgs boson in the particle spectrum of our model.  
	Two charged goldstone bosons are eaten up by the $W_L$ and $W_R$ gauge bosons to give them mass, while two
	neutral goldstone states provide mass to the $Z$ and $Z_R$.
	\section{Phenomenological Benchmark points} \label{sec:benchmarks}
	We now highlight some of the constraints on the parameter space in our model before discussing the
	choice of our benchmark points for the collider analysis.  
	We note that the most relevant constraints arise from the experimental bound on heavy gauge boson masses. 
	We also mention a few theoretical constraints that affect our parameter choices.
	
	\subsection{Theoretical and phenomenological constraints} 
	The theoretical constraints on some of the couplings in the model come from the requirement of perturbativity and unitarity. 
	The perturbativity condition requires that  at least at the electroweak symmetry breaking (EWSB) scale, all quartic couplings 
	satisfy
	\begin{equation}
		C_{H_i H_j H_k H_l} < 4\pi,
	\end{equation}
	whereas the Yukawa and gauge couplings need to be less than $ \sqrt{4\pi}$~\cite{Mohapatra}. Furthermore, tree-level 
	unitarity in the scattering of Higgs bosons and longitudinal components of EW gauge bosons necessitates that 
	the eigenvalues of the scattering matrices must be less than $16 \pi$~\cite{PhysRevLett.38.883, PhysRevD.16.1519}. 
	
	Additional constraints appear as a result of electroweak precision measurements, particularly the
	oblique parameters~\cite{PhysRevD.46.381}.
	The addition of extra Higgs doublets has little effect on the oblique parameter in general, particularly the $T$-parameter 
	because the custodial $SU(2)$ stays intact at the tree level. It can be broken at the loop level by the effect of other states. 
	However, the additional states in our model primarily belong to the right-handed sector and are rather heavy.
	Although the mixing between the left and right sectors is tiny, it cannot cause significant changes in the EW 
	$W$ or $Z$ boson masses. The consistency of our chosen benchmarks has been checked against existing limits~\cite{Jens}. 
	
	\subsection{Experimental constraints}
	The non-observation of any direct signal at the  LHC has put stringent constraints on the mass of the heavy 
	right-handed gauge bosons  $Z_R$, $W_R$ as well as the heavy Majorana neutrinos in LRSM. The different 
	constraints leading to limits on the model are briefly summarized below. 
	
	\paragraph{ Searches for heavy resonances:}
	\begin{figure*}
		\centering
		\includegraphics[width=0.49\textwidth]{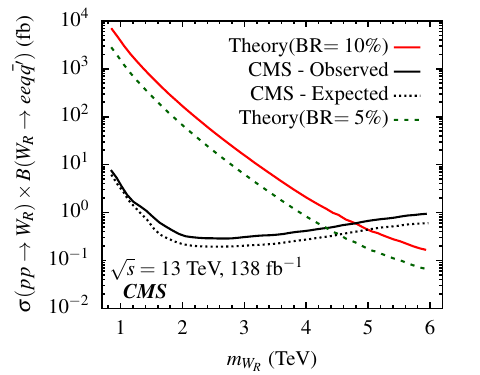}
		\includegraphics[width=0.49\textwidth]{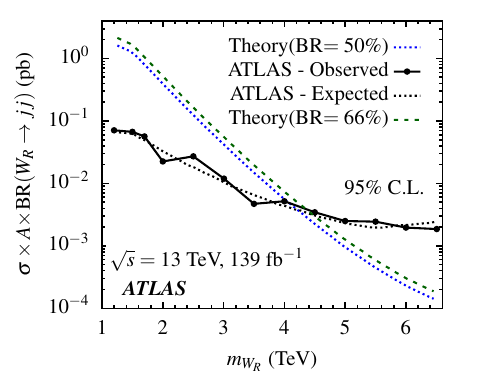}
		\caption{\label{fig:Br-ULR-MLR} Limit for the cross section times the branching ratio of $W_R$  as a function of its mass at $95\%$ C.L. are shown from  CMS heavy neutrino search~\cite{CMS:2021dzb} ({\em left-panel}) and ATLAS dijet search~\cite{ATLAS:2019fgd} ({\em right-panel}). Theoretical expectations for the cross sections are also shown with the branching ratios used in the experimental search ($10\%$ in {\em left-panel} and $50\%$ in {\em right panel} )	and in our analysis ($5\%$ in {\em left-panel} and $66\%$ in {\em right- panel} ). Please refer to the text for more details.}
	\end{figure*}
	The search for heavy resonances decaying to di-lepton/di-jet final state at the LHC put strong constraints on their 	production cross section.
	The major experimental constraint in our model comes from the $W_R$ search in the $l^\pm N$ channel ($N$ represents the heavy 
	right-handed neutrino) where the final state is either one charged lepton and a fatjet containing a high $P_T$ lepton~\cite{ATLAS:2019isd} 
	or two same-sign leptons and two jets~\cite{CMS:2018jxx,Kim:2018xib}, depending on the mass difference of $W_R$ and $N$. 
	The most recent	CMS search~\cite{CMS:2021dzb} puts a lower limit on $W_R$ which excludes its masses below $4.7$ TeV and 
	$5.0$ TeV for the electron ($e^\pm$) and muon ($\mu^{\pm}$) channels, respectively. 
	On the other hand, ATLAS~\cite{ATLAS:2019fgd} di-jet search  puts a lower bound of $4$ TeV on $W_R$ mass with SM-like couplings (i.e., $50\%$ branching in di-jet mode).  	
	These limits differ depending on the branchings of $W_R$  in the di-jet and the $l^\pm  N$ channels. 
	These dependencies are summarized in Fig.~6 of Ref.~\cite{CMS:2021dzb} and Fig. 4c of Ref.~\cite{ATLAS:2019fgd},  and one can extract the bound on $M_{W_R}$ 
	depending on the value of production cross sections times the branching ratios. 
	Both the CMS~\cite{CMS:2021dzb} and ATLAS~\cite{ATLAS:2019isd} collaborations have used the 
	minimal LRSM (MLRSM) model in their search for the heavy $W_R$ boson, where  the $W_R\to e N$ decay 
	branching ratio is about $10\%$ for large values of the $W_R$ mass~\cite{Mattelaer:2016ynf}. In our case, the 
	branching ratio is relatively smaller ($<5\%$) due to new modes of decay available for the $W_R$ boson into heavy 
	leptons (seesaw partners) and heavy neutrinos, as shown in Table~\ref{tab:branchings-BP1}. 
	As the decay branching ratio of $W_R\to e N$ is suppressed, limits on the $W_R$ mass are relaxed (see {\em left-panel} of Fig.~\ref{fig:Br-ULR-MLR})  when 
	compared to the CMS~\cite{CMS:2021dzb} $4.7$ TeV bound in the electron channel.
	In the di-jet mode, one notes that the branching for $W_R$ is $66\%$ in our model, and thus the limit on $W_R$ mass increases  (see {\em right-panel} of Fig.~\ref{fig:Br-ULR-MLR}). We have a common limit of $4.3$ TeV on $W_R$ mass coming either from $l^\pm N$ or di-jet search
	for our parameter choices.
	
	The heavy neutral gauge boson ($Z_R$) has a rather weak limit from the direct di-jet search~\cite{CMS:2018wxx} excluding $m_{Z_R}<2.9$ TeV having SM-like gauge couplings with	the SM fermions. However, in LRSM the $Z_R$ is heavier than $W_R$, and thus a strong limit on $W_R$ mass indirectly puts a stronger limit on $Z_R$ mass (see Eqs.~(\ref{eq:WR}) and (\ref{eq:ZR})). The $4.3$ TeV lower limit on $W_R$ mass, as discussed above, puts a lower limit of $5$ TeV on $Z_R$ mass.

	In addition, we have also considered heavy neutrino search results from the LHC which put limits on 
	$m_N$~\cite{ATLAS:2018dcj,CMS:2018agk,CMS:2018jxx}, and these are included in our benchmark selection. 
	
\paragraph{Bounds from FCNC and Higgs searches:} 
We avoid flavor-violating interactions of charged leptons with neutral scalars by choosing diagonal configurations for the corresponding Yukawa matrices involving the SM and heavy exotic leptons. There are no flavor-changing neutral Higgs (FCNH) interactions in our model at the 
 tree-level because of the diagonal couplings for the neutral scalars.  
Heavy non-standard neutral scalars, pseudoscalars and the charged Higgses have strong bounds~\cite{Aoki,PhysRevD.100.094508} from low energy flavour-changing neutral current (FCNC) effects, especially from the $K$-meson and $B$-meson mass mixing. These bounds have been studied for the MLRSM in Ref.~\cite{Yue}. As the FCNH effects are absent, these bounds do not apply on our neutral states. The $W_R$ mass limits coming from the meson mixings are much weaker than the direct collider bounds.

The signal intensity measurements of the $125$ GeV Higgs in several final states, including $ZZ$, $WW$, $bb$, $\tau \tau$, and $\gamma \gamma$ final states~\cite{Mcbrayer, Avramidou}, provide additional constraints, which have been considered while choosing our benchmark points. The gauge and Yukawa couplings of the $125$ GeV scalar seem to lie extremely close to their SM value, as suggested by the experimental results. We thus limit our study to the alignment limit, i.e., $|y^V_h| \sim 1$, which is the modification factor to the $hVV$ ($V=W/Z$) couplings due to new physics. The parameter space of interest is further restricted by collider searches for non-standard neutral scalar states and charged scalar states. These searches are carried out at the LHC~\cite{CMS:2021yci,ATLAS:2022eap,ATLAS:2022rws,ATLAS:2021upq,CMS:2022jqc} in a variety of SM final states limiting their masses in the range of $1$ to $2$ TeV. As our non-standard neutral and charged scalar states are considered to be very heavy they do not affect our analysis. 
	
\subsection{Chosen parameter space}
Motivated by the signal of fatjet searches arising from heavy vector-like leptons and heavy neutrinos, we choose the heavy 
charged lepton mass to be $\sim 0.72- 1$ TeV while three heavy neutrino masses in the range of $\sim 180$ --$700$ GeV and 
$m_{W_R} \approx 4.5-5$ TeV, which are consistent with all experimental observations. The remaining BSM particles, including 
heavy Higgs (charged and neutral), pseudoscalar, heavy quarks, and the remaining six heavy neutrinos, are kept at very high 
mass ($>5$ TeV), which does not contribute to the collider signatures considered in this article. We choose three benchmark points ({\tt BP}) having  $m_{W_R}=4.5$ in {\tt BP1}  and $m_{W_R}=5$ TeV in {\tt BP2} and {\tt BP3}. For {\tt BP3}, we further subdivide the benchmark through one heavy charged mass. The heavy charged leptons are kept around $1$ TeV in  {\tt BP1} and {\tt BP2} (low boost), while they are kept around $0.72$ TeV in {\tt BP3} (high boost). The scalar and gauge boson masses are shown in Table~\ref{tab:boson-mass} for the three benchmark points. In Table~\ref{tab:fermion-mass}, we show fermion masses chosen for our phenomenological study.
	\begin{table}
		\caption{\label{tab:boson-mass} Gauge boson and scalar masses  with the common input parameters   $\a_1 = -0.2,~\a_2=0.1,~\l_{11}=0.1307,~\l_{12}=0.8,~\l_{13}=0.05,~\l_{14}=-0.1,~\l_{22}=0.5,~\l_{23}=0.1,~\l_{24}=0.1,~\l_{33}=0.2,~\l_{34}=0.1,~\l_{44}=0.1,\mu_{12}^2=-2.5\times10^4,~\mu_{34}^2=-8.5\times10^7$, $v_{LQ}/\sqrt{2}=173.4$ GeV, $v_{Ll}/\sqrt{2}=14$ GeV. The right handed VEVs for {\tt BP1}  are $v_{RQ}=10.6$  and $v_{Rl}=8.4$ TeV, and that for {\tt BP2} and {\tt BP3} are $v_{RQ}=12.48$ TeV and $v_{Rl}=8.4$ TeV.  }	
		\centering
		\renewcommand{\arraystretch}{1.3}
		\begin{tabular*}{0.45\textwidth}{@{\extracolsep{\fill}}lll@{}}\hline
			Particle & 	\multicolumn{2}{c}{Mass} \\ \hline
			&  {\tt BP1} & {\tt BP2} and  {\tt BP3} \\ \hline
			$W_R$ & $4.5 $ TeV & $5.0 $ TeV\\\hline
			$Z_R$ &  $5.33 $ TeV & $5.93 $ TeV \\\hline
			$H_1$  & $125.8$ GeV & $126.0$ GeV \\\hline
			$H_2$  & $6.38 $ TeV & $7.4 $ TeV \\\hline
			$H_3$  & $7.51$ TeV & $ 8.15$ TeV \\\hline
			$H_4$  & $13.89 $ TeV & $14.31 $ TeV \\\hline
			$A_1$  & $7.51 $ TeV & $8.15 $ TeV \\\hline
			$A_2$  & $13.2 $ TeV & $13.52 $ TeV \\\hline
			$H_1^+$  & $7.51 $ TeV & $8.15 $ TeV \\\hline
			$H_2^+$  & $13.2$ TeV & $13.52 $ TeV \\ \hline
		\end{tabular*}
	\end{table}
 
	\begin{table*}[ht!]
		\caption{The heavy quarks, charged and neutral lepton masses for the benchmark points with the input parameters shown in Eqs.~(\ref{eq:heavy-fermion-matrix}) and (\ref{eq:yukawa}).}
		\label{tab:fermion-mass}	
		\centering
		\renewcommand{\arraystretch}{1.5}
		\begin{tabular*}{\textwidth}{@{\extracolsep{\fill}}llll@{}}\hline
			Up-type Quark&Down-type Quark &  Charged Lepton& ~~~~~~~~~~~~~Neutrino \\ 
			(TeV) & (TeV) & (TeV) & ~~~~~~~(GeV)~~~~~~~~~~~~~ (TeV) \\ \hline
			$\begin{array}{c}
				{\tt BP1}, {\tt BP2}, {\tt BP3}  \\\hline
				m_U =11.94 \\ m_C=11.96 \\m_T=30.0
			\end{array}
			$
			&
			$\begin{array}{c}
				{\tt BP1}, {\tt BP2}, {\tt BP3}  \\\hline
				m_D =16.0 \\ m_S=20.1 \\m_B=30.0
			\end{array}
			$
			&
			$\begin{array}{c|c}
				{\tt BP1}, {\tt BP2} & {\tt BP3}  \\\hline
				m_{E_4} =1.17 & 0.721 \\ m_{E_5}=2.0 & 1.3  \\m_{E_6}=2.56 & 1.87
			\end{array}
			$
			&
			
			\begin{tabular}{c|c}
				\multicolumn{2}{c}{{\tt BP1}, {\tt BP2}, {\tt BP3}}	 \\\hline		
				$m_{\nu_4} =186.0$ & $m_{\nu_7} =5.14$\\ $m_{\nu_5} =404.8$ &  $m_{\nu_8} =5.30$ \\ $m_{\nu_6} =689.7$ & $m_{\nu_9} =5.50$ \\
				& $m_{\nu_{10}} =15.05$\\ & $m_{\nu_{11}} =15.11$\\ & $m_{\nu_{12}} =15.19$\\
			\end{tabular}
			\\ \hline
		\end{tabular*}
	\end{table*}
 
	\begin{figure*}
		\centering
		\includegraphics[width=0.49\textwidth]{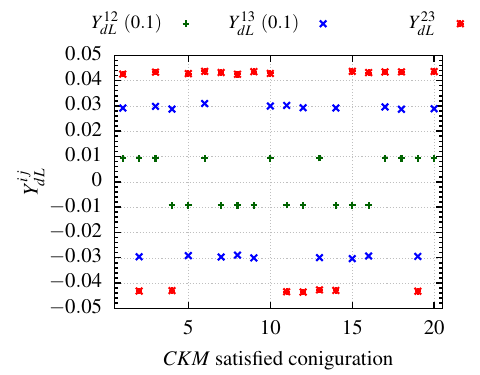}	
		\includegraphics[width=0.49\textwidth]{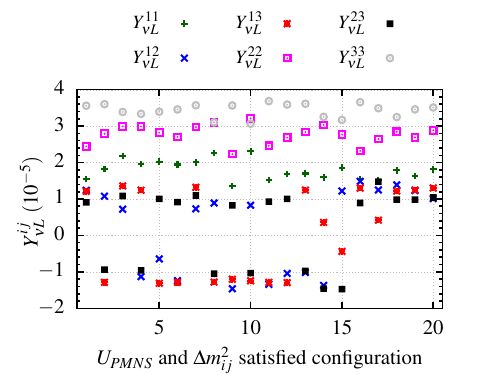}	
		\caption{\label{fig:ckm-parameter} The CKM satisfied points for the $Y_{dL}$, and the $U_{PMNS}$ and $\Delta m_{ij}^2$ satisfied points for the $Y_{\nu L}$ are shown in the {\em left-panel} and {\em right-panel}, respectively for {\tt BP1}.}
	\end{figure*}
	All the Yukawa matrices in Eq.~(\ref{eq:Yukawa}) are chosen diagonal except the $Y_{dL}$ and $Y_{\nu L}$ to generate experimentally measured CKM parameters and neutrino mixing matrix $U_{PMNS}$ as discussed before. The diagonal heavy singlet fermion (quark and lepton) mass matrices are taken to be
	\begin{eqnarray}\label{eq:heavy-fermion-matrix}
		M_U&=&{\text{Diag}}\left(30.0,11.93,6.6\right)\times 10^3,\nonumber\\
		M_D&=&{\text{Diag}}\left(1.6,2.0,3.0\right)\times 10^4,\nonumber\\
		M_E&=&{\text{Diag}}\left(0.8,1.0,1.0\right)\times 10^3,\nonumber\\
		M_R &=& M_L = {\text{Diag}}\left(1.0,1.0,1.0\right)\times 10^4,\nonumber\\
		M_N &=& {\text{Diag}}\left(1.0,1.0,1.0\right)\times 5\times 10^3
	\end{eqnarray}
	in units of GeV. 
	These inputs for the singlets with appropriate Yukawa coupling matrices produce the desired mass of SM fermions 
	together with the heavy fermion eigenstates shown in Table~\ref{tab:fermion-mass}. The Yukawa couplings are 
	chosen as
	\begin{align}\label{eq:yukawa}
		&Y_{u(L,R)}^{11}\sim Y_{d(L,R)}^{11}\sim Y_{e(L,R)}^{11} \simeq 10^{-2},\nonumber\\
		&Y_{u(L,R)}^{22}\sim Y_{dR}^{22}\sim Y_{e(L,R)}^{22} \simeq 10^{-1},\nonumber\\ 
		&Y_{u(L,R)}^{33}\sim Y_{dL}^{33}\sim Y_{eR}^{33} \simeq 1.0 .
	\end{align} 
We require $Y_{dR}^{33}\simeq0.092(0.078)$ in {\tt BP1} ({\tt BP2} and {\tt BP3}); $Y_{eL}^{33}\simeq0.146(0.132)$ in {\tt BP1} and {\tt BP2} ({\tt BP3}) to keep the third generation heavy fermion masses in few TeV range. We set the off-diagonal elements in $Y_{dL}$ matrix by fitting them to get the experimentally measured CKM mixing matrix~\cite{ParticleDataGroup:2020ssz} within $4\sigma$ error, with $Y_{dL}^{21}=Y_{dL}^{32}=Y_{dL}^{31}=0$. The values of $Y_{dL}^{12}$, $Y_{dL}^{13}$, and  $Y_{dL}^{23}$ are shown in Fig.~\ref{fig:ckm-parameter} in {\em left-panel} for different scanning configurations (horizontal axis) in {\tt BP1}. The values are $|Y_{dL}^{12}|\simeq 10^{-3}$, $|Y_{dL}^{13}|\simeq 3\times 10^{-3}$, and $|Y_{dL}^{23}|\simeq 4\times 10^{-2}$ for all scanning configuration. 
The values of $Y_{dL}^{12}$, $Y_{dL}^{13}$, and $Y_{dL}^{23}$  are of the same order of magnitude as above, for all our benchmark points.
	
Similar to the quark sector Yukawa, we keep the $Y_{\nu R}$ diagonal while $Y_{\nu L}$ is chosen non-diagonal in order to obtain the neutrino mixing matrix $U_{PMNS}$. We choose $Y_{\nu R}={\text{Diag}}\left(0.2,0.3,0.4 \right)$ and scan the parameter space for $Y_{\nu L}^{ij}$ which is shown in Fig.~\ref{fig:ckm-parameter} in the  {\em right-panel} with different scanned configurations (horizontal axis) which satisfy  $6.82\times 10^{-5}~\text{eV}^2<\Delta m_{21}^2< 8.04\times 10^{-5}~\text{eV}^2$, $2.43\times 10^{-3}~\text{eV}^2<\Delta m_{31}^2<2.60\times 10^{-3}~\text{eV}^2$  and $U_{PMNS}$  mixing elements within $3\sigma$ error~\cite{Esteban:2020cvm,NuFit50}. The relative values of $Y_{\nu L}^{ij}$ are below $4\times 10^{-5}$ keeping the relative order intact among themselves. We have chosen the first point in the scanned configuration in our analysis.

	\begin{table*}[ht!]
		\caption{\label{tab:branchings-BP1} The relevant branching ratios are listed for the decay channel of the heavy gauge bosons $W_R$ and $Z_R$,  heavy leptons $E_i$, and lightest three heavy neutrinos $\nu_i$, ($i=4,5,6$) in {\tt BP1}. }
		\centering	
		\renewcommand{\arraystretch}{1.4}
		\begin{tabular*}{0.3\textwidth}{@{\extracolsep{\fill}}|ll@{}}\hline
			Decay & Branching \\ \hline
			$W_R^+\to jj$ & 66.0\% \\
			$W_R^+\to t\bar{b}$ & 10.2\% \\
			$W_R^+\to e^+\nu_4$ &   5.03 \% \\
			$W_R^+\to \mu^+\nu_5$ &   1.58 \% \\
			$W_R^+\to \tau^+\nu_6$ &   2.43 \% \\
			$W_R^+\to E_4^+\nu_4$ &   5.16 \% \\
			$W_R^+\to E_5^+\nu_6$ &   5.08 \% \\
			$W_R^+\to E_6^+\nu_5$ &   4.60 \% \\\hline
		\end{tabular*}	
		\begin{tabular*}{0.3\textwidth}{@{\extracolsep{\fill}}|ll@{}}\hline
			Decay & Branching \\ \hline	
			$Z_R\to jj$ & 47.5\% \\  
			$Z_R\to \nu_4\nu_4$ & 7.17\% \\ 
			$Z_R\to \nu_5\nu_5$ & 6.56\% \\
			$Z_R\to E_4\bar{E_4}$ & 2.23\% \\  
			$Z_R\to E_5\bar{E_5}$ & 1.15\% \\  
			& \\
			& \\
			& \\ \hline
		\end{tabular*}	
		\begin{tabular*}{0.3\textwidth}{@{\extracolsep{\fill}}|ll|@{}}\hline
			Decay & Branching \\ \hline	
			$E_4\to \nu_4 jj$ & 70.6\% \\
			$E_5\to \nu_6 jj$ & 8.68\% \\
			$E_6\to \nu_5 jj$ & 73.7\% \\ 
			$E_6\to \mu jj$ & 3.65\% \\ 			
			$\nu_4\to e^\pm jj$ & 100\% \\
			$\nu_5\to \mu^\pm jj$ & 94.5\% \\
			$\nu_6\to \tau^\pm jj$ & 87.3\% \\ 
			& \\\hline
			
		\end{tabular*}	
		
	\end{table*}		
	
	\section{Collider Analysis}\label{sec:collider-analysis}		
	In this section, we study the collider signatures of our model at the high luminosity run of the LHC 
	with the chosen benchmark parameters discussed above. 
	Even though the exotic charged and neutral leptons are very heavy and have very limited production rates through SM 
	gauge bosons, they  
 can be resonantly produced through the heavy $SU(2)_R$ gauge bosons with comparatively large rates. 
	We therefore  focus on the production of heavy neutrinos and heavy charged leptons through the  $s$-channel processes mediated by the heavy right-handed gauge bosons. Once produced, these heavy charged leptons ($E_i$'s)  and neutrinos ($\nu_i$, where $i \ge 4$) finally decay to SM leptons and jets with almost $100\%$ branching ratios, as shown in Table~\ref{tab:branchings-BP1}  for {\tt BP1}. 
        The branching ratios for the other benchmark points, viz. {\tt BP2} and {\tt BP3}  are shown in Table \ref{tab:branchings-BP23} in the \ref{app:branching}.
 The decay of these heavy fermions (produced from the decay of very heavy $SU(2)_R$ gauge bosons) leads to 
	the interesting possibility of final states with boosted objects. 
What we finally observe are some clusters of highly boosted and collinear particles, which can only be detected as a large radius jet in the detector. 
The presence of these fatjets play an important role in achieving good signal significance over the estimated SM background. 
In addition, the fatjet from the decay of heavy charged lepton also contains a non-isolated SM charged lepton along with two sub-jets. 
 The fatjet originating from the heavy neutrino decay contains two sub-jets, as shown in Eq.~\ref{eq:3-fatjet-signal}. 
 In our work, we therefore look for fatjets containing charged leptons and analyze these fatjets by studying their {\it substructure}. 
We consider  signals with different fatjet multiplicities,  that include inclusive two, three, and four fatjets. Note that our primary production channel
involves particles ($E_i$ and $\nu_i$) that carry a lepton number which leads to some of the 
	fatjets having a non-isolated charged lepton ($e,\mu$) within the fatjet radius. We use the jet-substructure 
	techniques~\cite{Brust:2014gia} to identify the sub-jet associated with these boosted charged leptons  when they cannot be isolated at the detector.

	\subsection{Signal and background}  
	\begin{figure}
		\centering
		\includegraphics[width=0.49\textwidth]{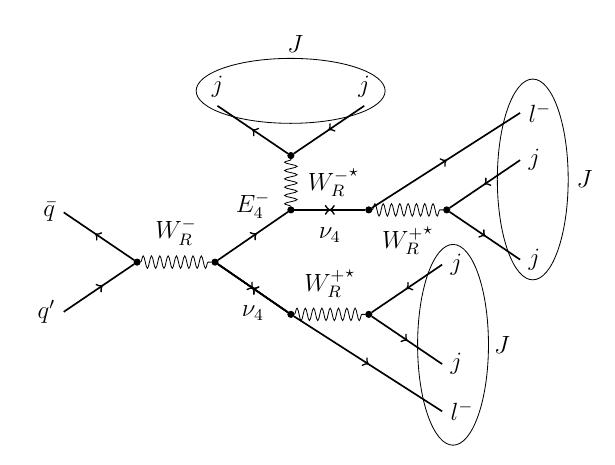}
		\caption{\label{fig:feynma-3J} Representative Feynman diagram for three fatjet signal.}	
	\end{figure}
	Here we describe the fatjet signals we are interested in, followed by the possible SM background in detail.
	We have generated events for all our signal and background processes using {\tt MadGraph5\_aMC@NLO} v2.7.3~\cite{Alwall:2014hca} at 
	leading order (LO) in QCD with a dynamic choice of factorization scale given by $\sum M_i^T/2$, where $M_i$ is the 
	transverse mass of final state particles.  We use {\tt nn23lo1}~\cite{NNPDF:2014otw} for the parton distribution 
	functions (PDFs). Events are passed to {\tt PYTHIA8}~\cite{Sjostrand:2014zea} for showering and 
	hadronization followed by a fast detector simulation in 
	{\tt Delphes v3.4.2}~\cite{deFavereau:2013fsa}, with added pile-up events embedded in {\tt Delphes v3.4.2}
	for the high luminosity LHC (HL-LHC). The final state hadrons with non-isolated high $P_T$ leptons are clustered using 
	the Cambridge-Aachen algorithm of {\tt FastJet}~\cite{Cacciari:2011ma, Cacciari:2005hq} with a jet radius 
	$R_0 = 0.8$ (AK8 fatjet). 
	We groom~\cite{Salam:2010nqg} the fatjets using  soft-drop method~\cite{Dasgupta:2013ihk,Larkoski:2014wba} to remove soft and wide angle radiations after some pre-selection cuts. We use the soft radiation fraction parameter $z_{cut}=0.1$ and the angular exponent parameter $\beta=0$ for the grooming~\cite{CMS:2021dzb}.	
	The groomed AK8 fatjets are required to have a threshold on the soft-drop mass ($m_{SD}$) as  $m_{SD}> 40$ GeV.	
	
	To identify the sub-jets associated with 
 leptons and leading hadrons we employ the N-subjettiness technique, 
where we use `OnePass General ET General KT Axes’ with $p = 0.6$ (for kt 
and Cambridge-Aachen axis choice of $p = 1$ and $p = 0$, respectively)~\cite{Catani:1993hr, Catani:1991hj}. To use this method for finding sub-jets out of a fatjet, we use {\tt Fastjet Contrib}~\cite{Thaler:2010tr}. The $t\bar{t}$ background was estimated at 
next-to-next LO (NNLO) using a $k$-factor of $1.6$~\cite{Catani:2019hip}.
	
\subsubsection{Fatjet Signals}\label{sec:signal}
We consider three types of signals containing fatjets for our analysis which are described below. Note that all the signal fatjets have $P_T >$ 200 GeV before grooming. This pre-selection cut on the reconstructed fatjets is
imposed to suppress the SM background, especially the large QCD multijet background. As we wish to study the signal which dominantly comes from 
the production of the heavy $SU(2)_R$ gauge bosons which have masses above $5$ TeV, the fatjets originating from them are expected to have significantly large $P_T$. This pre-selection helps in removing the background contributions without compromising on the signal events.
	\paragraph{Two fatjet inclusive searches:} In this signal topology, our final state contains at least 
	two high $P_T$ fatjets, with each 
 containing one high $P_T$ lepton lying inside the fatjet radius.  The five different production channels  that contribute to this  final state in our model are 
	\begin{enumerate}[i.]
		\item {\bf Signal $1$:} ~$ p p \to W_R,~ W_R \to E_4 \nu_4,~ E_4 \to \nu_4 j j$ (both the $\nu_4$ then decay into $e$/$\mu$ with additional two jets),
		\item {\bf Signal $2$:} ~$ p p \to W_R,~ W_R \to E_6 \nu_5,~ E_6 \to \nu_5 j j$ (both $\nu_5$ will then decay into $e$/$\mu$ with additional two jets),
		\item {\bf Signal $3$:} ~$ p p \to W_R,~ W_R \to E_6 \nu_5,~ E_6 \to \mu j j$ ($\nu_5$ then decays into $e$/$\mu$ with additional two jets),
		\item {\bf Signal $4$:} ~$ p p \to Z_R,~ Z_R \to (\nu_4 \nu_4)/(\nu_5 \nu_5)$ ($\nu_4/\nu_5$ then decay into $e$ and $\mu$ with additional two jets),
		\item {\bf Signal $5$:} ~$ p p \to Z_R,~ Z_R \to E_4 \bar{E_4},~ E_4 \to \nu_4 j j$ ($\nu_4$ then decays into $e$/$\mu$ with additional two jets).
	\end{enumerate}
 
	\paragraph{Three-fatjet inclusive searches:} Here our final state contains at least three high $P_T$ fatjets, where the two 
 leading ($P_T$-ordered) fatjets must contain one high $P_T$ lepton in them.  The three 
	production channels that dominantly contribute to this three-fatjet signal are given by
	\begin{enumerate}[i.]
		\item {\bf Signal $1$:} ~$ p p \to W_R,~ W_R \to E_4 \nu_4,~ E_4 \to \nu_4 j j$ (all $\nu_4$ then decay into  $e$/$\mu$  with additional two jets), a representative Feynman diagram for this channel is shown in Fig.~\ref{fig:feynma-3J},
		\item {\bf Signal $2$:} ~$ p p \to W_R,~ W_R \to E_6 \nu_5,~ E_6 \to \nu_5 j j$ (both the $\nu_5$ then decay into $e$/$\mu$ with additional two jets),
		\item {\bf Signal $3$:} ~$ p p \to Z_R,~ Z_R \to E_4 \bar{E_4},~ E_4 \to \nu_4 j j$ (each $\nu_4$ then decays into $e$/$\mu$ with additional two jets).
	\end{enumerate}
	
	\paragraph{Four-fatjet inclusive searches:} In this signal, the final state contains at least four high $P_T$ fatjets with 
	the leading two fatjets inclusive of only one high $P_T$ lepton within their jet radius. 
	For this case we find that only $ p p \to Z_R,~ Z_R \to E_4 \bar{E_4},~ E_4 \to \nu_4 j j$ contributes dominantly 
	where $\nu_4$ as before decays into $e/\mu + j j$. 

 All individual contributions in the respective final states are summed together and identified as the $signal$.
 
	\subsubsection{Backgrounds}
For the above signals, we consider the following SM background.
\paragraph{QCD multi-jet:} QCD multi-jet events will be a major source of background due to its large production cross section. We primarily focus on the $4$-jet final state, which can lead to our two, three, or four fatjet signals where some jet could also fake a lepton as a part of the fatjet. We however find that a large $P_T$ cut on the fatjet and a cut on the jet invariant mass  helps to reduce the QCD background by 
a significant amount.  
\paragraph{Top pair production ($t\bar{t}$):} The other dominant background  for our signal comes from the 
SM production of $t \bar{t}$ where the leptonic decay mode of 
$t \bar{t}$ can give us fatjets with two high $P_T$ lepton as part of the fatjets. On the other hand, the semi-leptonic and fully hadronic decay mode of $t \bar{t}$ can also be a dominant background where one or two hadrons can fake as leptons or have leptonic decays in the detector.
\paragraph{$W/Z$+jets and $tW$ :} The weakly produced $ZW$ and $tW$ processes can also be possible sources of fatjet background. However, the production cross section of both these processes are very small compared 
to the QCD and $t\bar{t}$ background. In addition, for the $tW$ case the probability 
	of getting at least two heavy and highly boosted fatjets is much smaller than $t \bar{t}$ due to the lower mass of $W$. The other SM subprocesses that can give fatjet signatures are $W/Z$+jets. 
	Although the production cross section for $W/Z$+jets is comparatively much larger than $tW$, it has 
	less probability of giving at least two high-energy leptons in  two separate fatjets, when compared to the $t \bar{t}$ leptonic decay case. 
 We therefore, neglect the above backgrounds in our analysis.

	\subsection{Variables}
	To achieve good signal significance over the aforementioned backgrounds, we 
 identify some kinematic variables  
	which have characteristically different distributions for the signal compared to the SM background. 
	After analyzing various distributions, we find the relevant variables to be the $P_T$ and invariant mass of the fatjets, $P_T$ of the sub-jets associated with leptons, and two interesting substructure variables $LSF$ and $LMD$. 
 
	The $P_T$-ordered jets and the jet-mass are denoted by ${P_T}_{j_i}$ and $m_{j_i}$ where $i=1,2,3,4$. 
 The invariant mass of any two, three, and four fatjets are also important variables and are called  $m_{{j_1}\ldots{j_n}}$. In addition, we include two jet substructure variables that have been used for fatjet analyses in CMS searches~\cite{CMS:2021dzb}, viz.
the lepton sub-jet fraction ($LSF$) and lepton mass drop ($LMD$) of the sub-jet associated with charged 
	leptons~\cite{Brust:2014gia}. 
	\begin{figure*}[!hptb]
		\centering
		\includegraphics[width=0.495\textwidth]{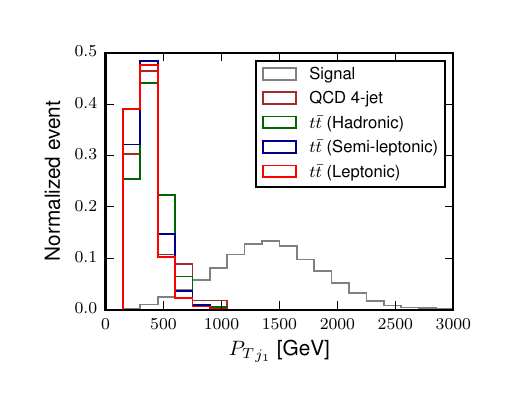}
		\includegraphics[width=0.495\textwidth]{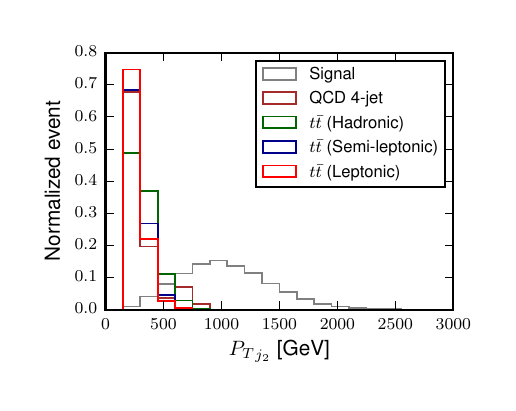} \\
		\includegraphics[width=0.495\textwidth]{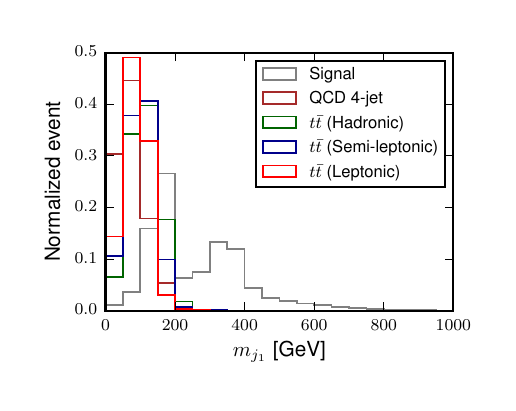}
		\includegraphics[width=0.495\textwidth]{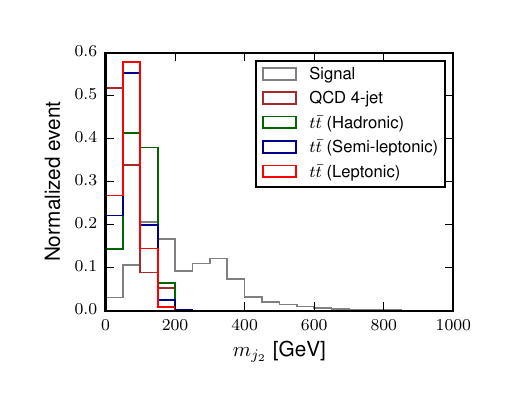} \\
		\includegraphics[width=0.495\textwidth]{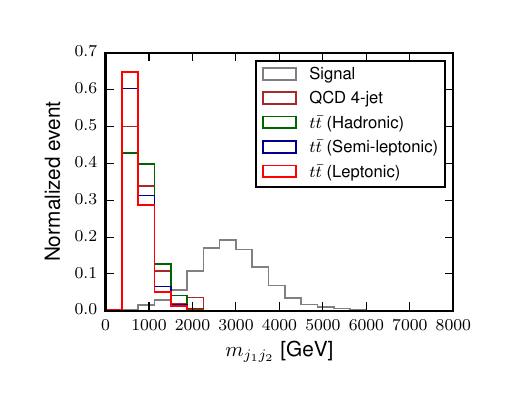} 
		\caption{Normalized distributions of the signal and different SM backgrounds as a function of the transverse momentum ($P_T$) of the two leading fatjets ({\em top-panel}), the mass ($m_j$) of two leading fatjets ({\em middle-panel}), and invariant mass ($m_{j_1 \, j_2}$) of the two leading fatjets ({\em bottom-panel}) in the {\it two-fatjet inclusive} searches for {\tt BP2}.}
		\label{dist_2jet}
	\end{figure*}
	%
	%
	\begin{figure*}[!hptb]
		\centering
		\includegraphics[width=0.495\textwidth]{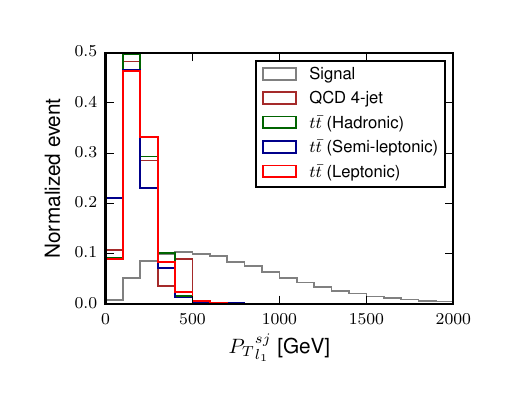}
		\includegraphics[width=0.495\textwidth]{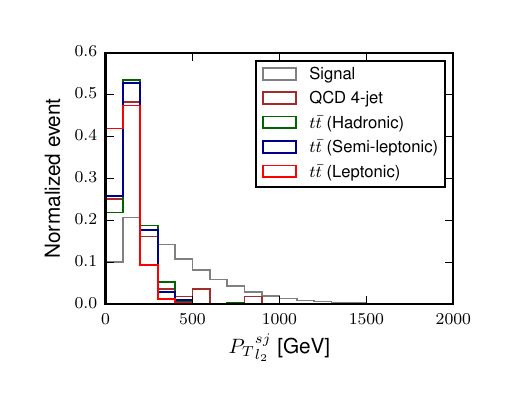}\\ 
		\includegraphics[width=0.495\textwidth]{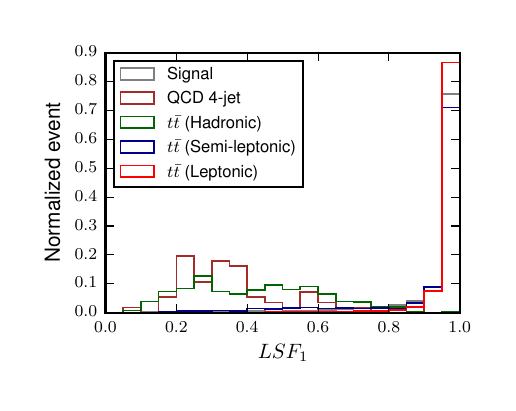}
		\includegraphics[width=0.495\textwidth]{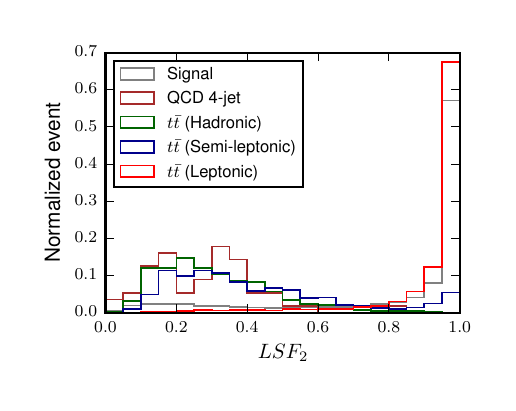} \\
		\includegraphics[width=0.495\textwidth]{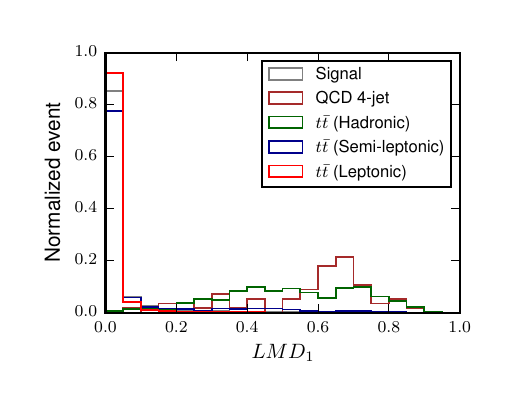} 
		\includegraphics[width=0.495\textwidth]{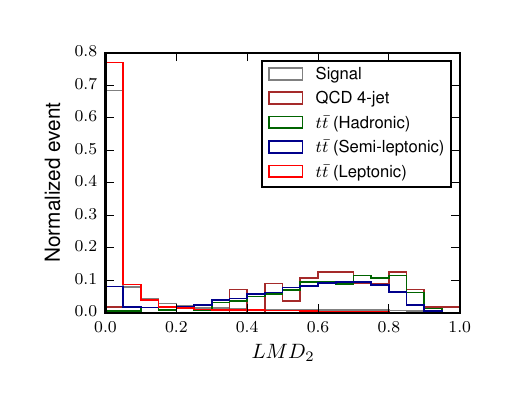}
		\caption{Normalized distributions of the signal and different SM backgrounds as a function of the transverse momentum of the leptonic sub-jet ($P_{T_{l}^{sj}}$) of the fatjets ({\em top-panel}), 
        the lepton sub-jet fraction variable involving the leading ($LSF_1$) and sub-leading ($LSF_2$) charged lepton ({\em middle-panel}) and the lepton mass drop variable with the leading ($LMD_1$) and sub-leading ($LMD_2$) charged lepton subtracted out ({\em bottom-panel}) in the {\it two-fatjet inclusive} searches for {\tt BP2}.}
		\label{dist1_2jet}
	\end{figure*}

	To compute the variables $LSF$ and $LMD$, we cluster all final state particles in the event (including leptons) into a fatjet. 
	For each fatjet we then use jet-substructure (JSS) algorithms to cluster its constituents in three sub-jets using N-subjettiness 
	techniques~\cite{Thaler:2010tr}.  We then check for sub-jets associated with all high $P_T$ leptons. Thereafter we calculate the lepton sub-jet fraction of each lepton (defined as the ratio between the lepton $P_T$ and $P_T$ of the sub-jet 
	associated with it) which is given by
	\begin{equation}
		LSF_n = \frac{{P_T}_{l_n}}{{P_T}^{sj}_{l_n}}
        \label{eq:lsf}
	\end{equation}
	where $n=1,2$. $LSF_1$ is associated with the leading lepton while $LSF_2$ is for the 
 sub-leading one. The lepton mass drop parameter is constructed in a similar way and given by
	\begin{equation}
		LMD_n = \frac{m^2_{sj-l_n}}{m^2_{sj}}
        \label{eq:lmd}
        \end{equation}
	where $m_{sj}$ represents the invariant mass of the sub-jet associated with the lepton and $m_{sj-l_n}$ represents the invariant mass of the same sub-jet with the $n^{th}$ lepton subtracted out. 
	
We now concentrate on the distributions of the aforementioned variables for two and three inclusive fatjet searches, where the 
signal and dominant SM background contributions are shown together to characterize the differences between their 
distributions. 
The normalized distributions are shown with events that have passed the following selection cuts:
\begin{align}
P_T^{\rm lepton} > 25 ~{\rm GeV}; && P_T^{\rm fatjet} > 200~ {\rm GeV}; && |\eta^{\rm fatjet}| <  4 .
\end{align}	
In Fig.~\ref{dist_2jet} and ~\ref{dist1_2jet}, we show the relevant distributions of the selected kinematic variables 
in the {\it two-fatjet 
inclusive} search for {\tt BP2}. 
In the {\em top-panel} of Fig.~\ref{dist_2jet}, we have shown the $P_T$ distribution of the leading and subleading fatjet where a pre-selection cut of $P_T >$ 200 GeV (before grooming for two leading fatjets) is in place. The distribution for the signal events peak at higher values of $P_T$ because of the large mass gap between the decay products and their mother particles. In contrast, the background events, say from the $t \bar{t}$ production where the jets come from the decay of SM $W$ bosons (including the $b$-jet), have relatively softer $P_T$ beyond the pre-selection cut. Note that for the top quarks to be highly boosted, they would have to be produced at large $\sqrt{\hat{s}}$ which would lead to a smaller cross section for the $t \bar{t}$ production. The QCD multijet background also falls off rapidly for large $P_T$. The {\em middle-panel} of Fig.~\ref{dist_2jet}  shows the jet mass ($m_j$) distributions of the two leading fatjets, while the {\em bottom-panel} shows the distribution of the invariant mass of the two leading fatjets. 
 For the signal events, one finds that the distributions in both $m_{j_1}$ and $m_{j_2}$ show two peaks that appear at the mass values of the heavy neutrinos (especially $\nu_4$ and $\nu_5$), whose highly boosted decay products correspond to the reconstructed fatjets. The jet-mass variable therefore helps us reconstruct the intermediate particles for the signal, which in turn also helps us to reject a large part of the SM background, where no such peak is observed at large values of $m_j$'s.
 
 Fig.~\ref{dist1_2jet} highlights the relevance of the substructure variables used in our analysis. The {\em top-panel} shows the event distribution of the signal and background as a function of $P_T$ of the 
 sub-jets inclusive of the leading and sub-leading charged lepton. As 
 expected, the signal fatjets which carry much larger $P_T$ than the fatjets in the SM background events also give larger $P_T$ sub-jets within them,   leading to the sub-jets also getting a hard $P_T$ distribution compared to the background. 
The {\em middle-} and {\em bottom-panels} of Fig.~\ref{dist1_2jet} display the event distributions as a function of $LSF$ and $LMD$ variables, respectively. The $LSF$ variables are found to sharply peak around $1$ for the signal and the $t\bar{t}$ (leptonic and semileptonic) background. This happens when the 
lepton is boosted enough and becomes one of the hardest constituents in the fatjet. As the signal and $t\bar{t}$ (leptonic and semileptonic) channels are most likely to give a fatjet with a non-isolated charged lepton within, these modes are the only ones that show the peak behavior around $1$. Note that the peak has vanished in the distribution of $LSF_2$  for the $t\bar{t}$ (semileptonic), as only one top decays in the leptonic channel. It is therefore less likely to give a second charged lepton inclusive fatjet. In addition, the charge leptons do not suffer from soft radiations into the sub-jets, and we can reconstruct the energy of the lepton by finding the sub-jet. In the case of the QCD-multijet and $t\bar{t}$ hadronic background, most of the events correspond to some jet faking a lepton. As they shower, they are surrounded by soft collinear radiation, and when we get the sub-jet associated with it, additional soft radiation gets added up. Thus the $LSF$ in most of the cases has a smaller value than one. The lepton mass drop ($LMD$) variable displays similar properties for the participating leptons within the fatjet. As the numerator in $LMD$, given by Eq.~\ref{eq:lmd} represents the invariant mass of the sub-jet with the charged lepton removed, it peaks near zero due to the unlikeliness of the absence of charged leptons  
in the $t \bar{t}$ (leptonic and semileptonic) and the signal events, whereas it gives non-zero values for QCD-multijet and $t\bar{t}$ hadronic background where the leptons are conspicuous by their absence. 

The kinematic distributions in the same variables for the three-fatjet inclusive analysis, show very similar features and are collected in  \ref{app:dist} as Figs.~\ref{dist_3jet}, \ref{dist1_3jet} and  \ref{dist2_3jet} for reference. The four-fatjet inclusive search is expected to be almost without any SM background and we focus only on the event rates 
for this final state.

	\begin{table*}[!h]
		\caption{ Cut summary for two-fatjet inclusive searches.}
		\label{tab:cutflow2jet_summary}
		\centering
		\renewcommand{\arraystretch}{1.5}
		\begin{tabular*}{\textwidth}{@{\extracolsep{\fill}}lllll@{}}\hline
			 &Cut summary & \\ \hline
		   {\bf Pre-selection cut} & {\bf Cut-A} & {\bf Cut-B} \\
			\hline
			
  $\bullet$ \,\, Number of leptons, $N_l \geq$ 2  with $P_T > 25$ GeV 
& $\bullet$ \,\, ${P_T}_{j_1} > 750$ GeV, ${P_T}_{j_2} > 500$ GeV  
& $\bullet$ \,\, ${P_T}_{j_1} > 750$ GeV, ${P_T}_{j_2} > 500$ GeV    \\

  $\bullet$ \,\,	Two fatjets with radius $R = 0.8$   
& $\bullet$ \,\, $LSF_1 > 0.9$, $LSF_2 > 0.9$ 
& $\bullet$ \,\, $LSF_1 > 0.9$, $LSF_2 > 0.9$\\

  $\bullet$ \,\,	Rapidity gap of two ungroomed fatjet, $|\Delta \eta| < 4$   
& $\bullet$ \,\, $LMD_1 < 0.05$, $LMD_2 < 0.05$ 
& $\bullet$ \,\, $LMD_1 < 0.05$,  $LMD_2 < 0.05$ \\
			 
& $\bullet$ \,\, ${P_T}^{sj}_{l_1} > 400$ GeV, ${P_T}^{sj}_{l_2} > 250$ GeV 
& \\

& $\bullet$ \,\, $m_{j_i} > 125$ GeV ($i=1,2$); $m_{j_1 j_2} > 1.5$ TeV & \\

			\hline
		\end{tabular*}
	\end{table*}
\subsection{Results}
	We now use the kinematic variables discussed in the previous section and exploit their characteristic distributions for signal and 
    background to define the event selection cuts we use to analyze 
    the three different fatjet signals. We have used a pre-selection criterion for signal and background in all our subsequent analyses 
    given by:

\medskip
	\noindent
	{\bf Pre-selection criteria :} For the pre-selection of events, we demand  at least two charged leptons ($e, \mu$) 
	with $P_T > 25$ GeV  at the event generation level along with two fat jets with radius $R = 0.8$ 
	and minimum $P_T > 200$ GeV which must lie within the rapidity gap of $|\eta| <  4$ for the ungroomed jet. As pointed out earlier, the strong pre-selection cut on the fatjet $P_T$ helps suppress the background events, dominantly coming from the QCD multijet processes. As the QCD background has a huge cross section, it is quite challenging to generate uniformly populated events in our simulations. The 
    pre-selection cuts at the generation level allow us to lower the large QCD cross section to manageable numbers for event generation. We have checked that the signal events are not affected much by the pre-selection cuts used in our analysis. 

	\begin{table*}[!h]
		\caption{ Signal and background events surviving after applying the pre-selection criteria as well as selection cuts at LHC with center-of-mass energy, $\sqrt s = 14$ TeV and ${\cal L} = 3000$  fb$^{-1}$ in the inclusive two-fatjet analysis for {\tt BP1}.}
		\label{tab:cutflow2jet_bp1}
		\centering
		\renewcommand{\arraystretch}{1.2}
		\begin{tabular*}{\textwidth}{@{\extracolsep{\fill}}lllll@{}}\hline
			Data sets&Cross section (pb)&\multicolumn{3}{c}{Number of events}\\ \hline
			& Parton-level & Pre-selection cut & {\bf Cut-A} & {\bf Cut-B} \\
			\hline
			
			$W_R \to e_6 \nu_5$, $e_6 \to \mu j j$  & $0.021\times 10^{-6}$ & 3 & 1 & 2   \\
			\hline
			$W_R \to e_6 \nu_5$, $e_6 \to \nu_5 j j$  & $0.400 \times 10^{-3}$ & 269 & 77 & 150   \\
			\hline
			$W_R \to E_4 \nu_4$ & $0.485 \times 10^{-3}$ & 660 & 51 & 225   \\
			\hline
			$Z_R \to (\nu_4 \nu_4)/(\nu_5 \nu_5)$  & $0.110 \times 10^{-3}$ & 189 & 40 & 93 \\
			\hline
			$Z_R \to E_4 \bar{E_4}$ & $0.012 \times 10^{-3}$ & 8 & 1 & 2 \\
			\hline				
			Total Signal ($s$)&& $1129$ & $170$ & $472$ \\ \hline
			QCD $4$-jet  & $90387.504$ & $1.52\times 10^7$ & $0$ & $0$\\
			\hline
			$t \bar{t}$ (Hadronic) & $229.603$ & $2.65\times 10^5$ & $0$ & $0$\\
			\hline
			$t \bar{t}$ (Semi-leptonic) & $178.569$ & $7.94\times 10^5$ & $0$ & $0$\\
			\hline
			$t \bar{t}$ (Leptonic) &  $34.854$ & $7.43\times 10^5$ & $105$ & $1.568\times 10^3$\\
			\hline
			Total Background ($b$) & & $1.7\times 10^7 $& $105 $& $1.568\times 10^3 $ \\ \hline
		\end{tabular*}
	\end{table*}
\paragraph{Two-fatjet inclusive searches:}
	The two-fatjet signal gives the largest cross section among all the fatjet signals. The signal  and background distributions shown in 
 Fig.~\ref{dist_2jet} and Fig.~\ref{dist1_2jet} are used to define the 
 event selection criterion to help achieve a good signal significance. We classify the kinematic selections for the signal and background events into two categories, which we call {\bf Cut-A} and {\bf Cut-B}.

	\medskip
	\noindent
	{\bf Cut-A} : 
\begin{enumerate}[i.]
		\item The  leading groomed fatjet is required to have ${P_T}_{j_1} > 750$ GeV, while  the sub-leading groomed fatjet
		should have ${P_T}_{j_2} > 500$ GeV.
		\item The jet mass of both the leading and sub-leading groomed fatjet should have a minimum value of $m_{j_i} > 125$ GeV ($i=1,2$) 
		while their  
		invariant mass must satisfy $m_{j_1 j_2} > 1.5$ TeV. 
		\item The  sub-jets associated with the leading charged lepton must have ${P_T}^{sj}_{l_1} > 400$ GeV 
		and sub-jets associated with the sub-leading charged lepton are required to have ${P_T}^{sj}_{l_2} > 250$ GeV. 
		\item The variables $LSF_1$ and $LSF_2$ need to be $> 0.9$, while $LMD_1$ and $LMD_2$ has an upper 
		limit satisfying $< 0.05$ after applying the grooming method. 
\end{enumerate}
	\bigskip
	\noindent
	{\bf Cut-B} : 
\begin{enumerate}[i.]
		\item The leading groomed fatjet is required to have ${P_T}_{j_1} > 750$ GeV, while  the sub-leading groomed fatjet 
		must have  ${P_T}_{j_2} > 500$ GeV. 
		\item The minimum value for the variables $LSF_1$ and $LSF_2$ is $> 0.9$,  while the 
		maximum value for  $LMD_1$ and $LMD_2$ is $< 0.05$ after applying grooming method. 
	\end{enumerate}
 A summary of all cuts mentioned here is also described in Tables~\ref{tab:cutflow2jet_summary}.
 
Note that the main difference between {\bf Cut-A} and {\bf Cut-B} is simply on the way the lepton inclusive fatjets are treated, 
	along with how we utilize the jet mass in the analysis which carries a bias of the mass of the new exotics. We list the number of signal and background events expected with an integrated luminosity of ${\cal L}=3000$ fb$^{-1}$ for {\tt BP1} in Table~\ref{tab:cutflow2jet_bp1}. We also list the cross sections for different signal sub-processes contributing to the two-fatjet inclusive search. The events with {\bf Cut-A} and {\bf Cut-B} are shown in 
Tables~\ref{tab:cutflow2jet_bp2} and \ref{tab:cutflow2jet_bp3} for  {\tt BP2},  and {\tt BP3} in \ref{app:tables}.

  As the signal events are peaked at very high $P_T$ (Fig.~\ref{dist_2jet}) the strong cut on their $P_T$ removes all of the background events coming from the QCD multijet and the hadronic decay of the $t \bar{t}$ production. In addition, the choice on the jet mass, invariant mass of the di-fatjet system and the 
  additional cuts on the jet substructure variables are used to exploit the presence of the highly boosted charged lepton and are helpful to remove the background events coming from the leptonic decay modes of the $t \bar{t}$ production (Fig.~\ref{dist1_2jet}).  
  For all the {\tt BP}s, we conclude that it is possible to remove the SM background with moderate efficiency using the pre-selection cuts, which is significantly improved after using {\bf Cut-A} and {\bf Cut-B}. We also note that {\bf Cut-A} is a much stronger selection criterion compared to {\bf Cut-B} as it requires an additional condition on the jet mass and invariant mass of the fatjet pairs, which reduces the signal events too. Therefore, {\bf Cut-A} will yield better significance where the 
  production cross section of the signal is relatively large, while {\bf Cut-B} will perform better where the signal cross section is small, and we do not want a signification reduction in the event rates for the signal. 
	
We calculate the signal significance~\cite{Cowan:2010js} using the formula
\begin{align}
	{\cal S} = \sqrt{2\left[(s+b) \log\left(1+\frac{s}{b}\right)-s\right]},
\end{align}
where $s$ and $b$ stand for the total number of signal and background events surviving after cuts and $\cal L$
is the integrated luminosity.  The signal significances for all the  {\tt BP}s in the inclusive two-fatjet final state are 
summarized in Table~\ref{tab:significance2jet}, with integrated luminosities of $3000$ fb$^{-1}$, $600$ fb$^{-1}$ and $300$ fb$^{-1}$ at the LHC with a center-of-mass energy $\sqrt{s}=14$ TeV.
	For {\tt BP1}, we are able to achieve a significance of more than $5 \sigma$ at ${\cal L} = 600$  fb$^{-1}$,  
	while for {\tt BP2}, which represents a point with a heavier $W_R$, we need $3000$  fb$^{-1}$ luminosity to achieve 
	$5 \sigma$ significance. This is due to the large mass of $W_R$ and $Z_R$ in {\tt BP2} compared to {\tt BP1}, which
	reduces the production cross section of the contributing sub-processes for {\tt BP2}, although the cut efficiency for the 
	signal events is better in the case of {\tt BP2}. In the case of {\tt BP3}, the signal significance is better compared to {\tt BP2} because of the larger production cross section and higher boost for the heavy leptons. For {\tt BP3}, an integrated luminosity of $600$ fb$^{-1}$ is required for a $5\sigma$ discovery.
	We note that  {\bf Cut-A} and {\bf Cut-B}  provide roughly the same amount of significance for {\tt BP1}. While for {\tt BP2} and {\tt BP3}, {\bf Cut-A} gives better significance as the jet mass and $P_T$ of sub-jets play an important role in reducing the background further.

	\begin{table}[!h]
		\caption{Signal significance for the two-fatjet signals for different cuts at the LHC with a center-of-mass energy $\sqrt{s}=14$ TeV with ${\cal L} = 3000$  fb$^{-1}$, ${\cal L} = 600$  fb$^{-1}$ and ${\cal L} = 300$  fb$^{-1}$.}
		\label{tab:significance2jet}		
		\centering
		\renewcommand{\arraystretch}{1.5}
		\begin{tabular*}{0.45\textwidth}{@{\extracolsep{\fill}}llll@{}}\hline
			After applying Cut & $3000$  fb$^{-1}$  & $600$  fb$^{-1}$ & $300$  fb$^{-1}$\\
			\hline
			{\bf Cut-A} ({\tt BP1}) & 13.77 $\sigma$ & 6.16 $\sigma$ & 4.35 $\sigma$ \\
			\hline
			{\bf Cut-B} ({\tt BP1}) & 11.39 $\sigma$ & 5.09 $\sigma$ & 3.60 $\sigma$\\
			\hline
			{\bf Cut-A} ({\tt BP2}) & $7.36$ $\sigma$ & $3.29$ $\sigma$ & $2.33$ $\sigma$\\
			\hline
			{\bf Cut-B} ({\tt BP2}) & $5.24$ $\sigma$ & $2.34$ $\sigma$ & $1.66$ $\sigma$\\
			\hline
			{\bf Cut-A} ({\tt BP3}) & 11.11 $\sigma$ & 4.97$\sigma$ & 3.51 $\sigma$ \\
			\hline
			{\bf Cut-B} ({\tt BP3}) & 7.64 $\sigma$ & 3.42 $\sigma$ & 2.42 $\sigma$\\
			\hline	
		\end{tabular*}

	\end{table}	
	
	\paragraph{Three-fatjet inclusive searches:} We have already discussed the major signal contribution to three-fatjet inclusive 
	searches for our scenario in Section~\ref{sec:signal}. The distributions of various variables are also shown in 
	Fig.~\ref{dist1_3jet} and Fig.~\ref{dist2_3jet}. To analyze the 
        three-fatjet final state events, we consider three different 
	sets of kinematic selections  ({\bf Cut-A}, {\bf Cut-B} and {\bf Cut-C}). A summary of all cuts mentioned below is also summarized in Tables~\ref{tab:cutflow3jet_summary}. 
	
	\medskip
	\noindent
	\noindent
	{\bf Cut-A} : 
\begin{enumerate}[i.]
		\item The leading groomed fatjet is required to have ${P_T}_{j_1} > 500$ GeV, while  the sub-leading groomed fatjet
		should have ${P_T}_{j_2} > 300$ GeV. 
		\item The jet mass of both the leading and sub-leading groomed fatjet should have a minimum value of $m_{j_i} > 150$ GeV ($i=1,2$),  
		while their invariant mass must satisfy $m_{j_1 j_2} > 800$ GeV. Additionally, the next-to-sub-leading groomed fatjet has jet 
		mass of $m_{j_3} > 100$ GeV.
		\item The  sub-jets associated with leading charged lepton must have ${P_T}^{sj}_{l_1} > 100$ GeV.
	\end{enumerate}
	
	\bigskip
	\noindent
	{\bf Cut-B} :
\begin{enumerate}[i.]
		\item The leading groomed fatjet is required to have ${P_T}_{j_1} > 750$ GeV, while  the groomed sub-leading fatjet
		should have ${P_T}_{j_2} > 500$ GeV. 
		\item The variables $LSF_1$ and $LSF_2$ have a minimum value of $ 0.9$,  while $LMD_1$ and $LMD_2$ 
		are bounded from above by $< 0.05$ after applying grooming techniques. 
	\end{enumerate}
	
	\bigskip
	\noindent
	{\bf Cut-C} : 
\begin{enumerate}[i.]
		\item The leading groomed fatjet is required to have ${P_T}_{j_1} > 750$ GeV, while  the sub-leading groomed fatjet
		should have ${P_T}_{j_2} > 500$ GeV. 
		\item The jet mass of both the leading and sub-leading groomed fatjet should have a minimum value of $m_{j_i} > 150$ GeV ($i=1,2$),  
		while their invariant mass must satisfy $m_{j_1 j_2} > 1000$ GeV. Additionally, the next-to-sub-leading groomed fatjet has jet 
		mass of $m_{j_3} > 100$ GeV.
		\item The  sub-jets associated with leading charged lepton must have ${P_T}^{sj}_{l_1} > 100$ GeV.
	\end{enumerate}

	\begin{table*}[!h]
		\caption{ Cut summary for three-fatjet inclusive searches.}
		\label{tab:cutflow3jet_summary}
		\centering
		\renewcommand{\arraystretch}{1.5}
		\begin{tabular*}{\textwidth}{@{\extracolsep{\fill}}lllll@{}}\hline
			 &Cut summary & \\ \hline
{\bf Pre-selection cut} & {\bf Cut-A} & {\bf Cut-B} & {\bf Cut-C} \\
			\hline
			
  $\bullet$ \,\, $N_l \geq$ 2 with $P_T > 25$ GeV  
& $\bullet$ \,\, ${P_T}_{j_1 (j_2)} > 500 (300)$ GeV 
& $\bullet$ \,\, ${P_T}_{j_1 (j_2)} > 750 (500)$ GeV
& $\bullet$ \,\, ${P_T}_{j_1 (j_2)} > 750 (500)$ GeV \\  
  $\bullet$ \,\, Two fatjets with radius $R = 0.8$.  
& $\bullet$ \,\, $m_{j_i} > 150$ GeV ($i=1,2$),
& $\bullet$ \,\, $LSF_1 > 0.9$, $LSF_2 > 0.9$
& $\bullet$ \,\, $m_{j_i} > 150$ GeV ($i=1,2$), \\
               & $m_{j_3} > 100$ GeV, $m_{j_1 j_2} > 800$ GeV 
&$\bullet$ \,\,  $LMD_1 < 0.05$, $LMD_2 < 0.05$ 
               & $m_{j_3} > 100$ GeV, $m_{j_1 j_2} > 1000$ GeV  \\	
  $\bullet$ \,\, $|\Delta \eta| < 4$ of two ungroomed fatjet 
& $\bullet$ \,\, ${P_T}^{sj}_{l_1} > 100$ GeV 
&  
& $\bullet$ \,\, ${P_T}^{sj}_{l_1} > 100$ GeV. \\
			\hline
		\end{tabular*}
	\end{table*}

	\begin{table}[!hb]
		\caption{Signal significance for the three-fatjet signals for different cuts at the LHC with a center-of-mass energy $\sqrt{s}=14$ TeV with ${\cal L} = 3000$  fb$^{-1}$, ${\cal L} = 600$  fb$^{-1}$ and ${\cal L} = 300$  fb$^{-1}$.}
		\label{significance3jet}	
		\centering
		\renewcommand{\arraystretch}{1.5}
		\begin{tabular*}{0.45\textwidth}{@{\extracolsep{\fill}}llll@{}}\hline
			After applying Cut & $3000$  fb$^{-1}$  & $600$  fb$^{-1}$ & $300$  fb$^{-1}$\\
			\hline
			{\bf Cut-A} ({\tt BP1}) & 3.13 $\sigma$ & 1.40 $\sigma$ & 0.99 $\sigma$\\
			\hline
			{\bf Cut-B} ({\tt BP1}) & 15.35 $\sigma$ & 6.87 $\sigma$ & 4.85 $\sigma$ \\
			\hline
			{\bf Cut-C} ({\tt BP1}) & 3.00 $\sigma$ & 1.34 $\sigma$ & 0.95 $\sigma$ \\
			\hline
			{\bf Cut-A} ({\tt BP2}) & $1.58$ $\sigma$ & $0.71$ $\sigma$& $0.50$ $\sigma$\\
			\hline
			{\bf Cut-B} ({\tt BP2}) & $7.92$ $\sigma$ & $3.54$ $\sigma$& $2.5$ $\sigma$\\
			\hline
			{\bf Cut-C} ({\tt BP2}) & $1.54$ $\sigma$ & $0.69$ $\sigma$& $0.49$ $\sigma$\\
			\hline
			{\bf Cut-A} ({\tt BP3}) & 2.08 $\sigma$ & 0.93 $\sigma$ & 0.66 $\sigma$\\
			\hline
			{\bf Cut-B} ({\tt BP3}) & 8.31 $\sigma$ & 3.72 $\sigma$ & 2.63 $\sigma$ \\
			\hline
			{\bf Cut-C} ({\tt BP3}) & 2.04 $\sigma$ & 0.91 $\sigma$ & 0.65 $\sigma$ \\
			\hline
			
		\end{tabular*}

	\end{table}

Note that the role of the cuts exploiting the $P_T$, jet mass, invariant mass of fatjet pairs, and the substructure variables, $LSF$ and $LMD$ remain very similar to the two-fatjet analysis, albeit with differences in the threshold choices for the event selection. 	
The effect of the three sets of cuts on our signal and SM background is tabulated in Table~\ref{tablecutflow3jet_bp1},  \ref{tablecutflow3jet_bp2}, and \ref{tablecutflow3jet_bp3} in \ref{app:tables} for {\tt BP1}, {\tt BP2},  and {\tt BP3}, respectively.
	Here, {\bf Cut-C} represents the strongest selection criteria for the events as it includes the additional requirements on the jet mass and a stronger cut on the di-fatjet invariant mass. We note that {\bf Cut-B} 
    which is the only set that includes the cut on the leptonic sub-jet variables, is useful in suppressing the  
    the SM background most and also retains the maximum signal events. This 
    is in agreement with our earlier observation in the two-fatjet analysis where we found that the jet-mass and invariant mass selections also suppress the signal events considerably. 
	The signal significances for all the {\tt BP}s are shown in Table~\ref{significance3jet} for ${\cal L} =  3000$  fb$^{-1}$, 
	${\cal L} = 600$  fb$^{-1}$, and $300$ fb$^{-1}$. 
 We find that {\bf Cut-B} that includes the $LSF$ and $LMD$ substructure variables, gives the best signal significance as it retains the maximum signal events for all the {\tt BP}s. 
 {\bf Cut-A} and {\bf Cut-C} give similar but lower significance. 
In the three-fatjet inclusive analysis, we require ${\cal L} = 600$  fb$^{-1}$ for {\tt BP1} and ${\cal L} \approxeq 3000$  fb$^{-1}$ for 
	{\tt BP2} and {\tt BP3} to achieve $5\sigma$ discovery.

	\paragraph{Four-fatjet inclusive searches:} 
	An interesting and more exotic signal for the model would be in the form of four fatjets. However, for such a final state, one can expect a very low signal cross section for our benchmark choices as only the $p \, p \to Z_R \to E_4 \bar{E_4}$ process
	contributes and is suppressed due to the large $Z_R$ mass (already bounded from below due to the
	$W_R$ mass limits in an LRSM framework). A part of the parameter space can be tuned to 
	generate a conducive set of masses to improve some decay branching fractions, but we are still constrained by the luminosity
	reach of the LHC. The SM background is expected to be quite negligible for such a final state and this signal may prove useful to consider at more high-energy machines such as the future $100$ TeV hadron collider~\cite{FCC:2018vvp}. As the LRSM gauge boson masses
    are excluded to be heavier and go beyond the reach of the LHC, the future hadron collider with $\sqrt{s}=100$ TeV will be able to produce them easily and will give significantly large cross sections~\cite{Nemevsek:2023hwx}. The large production rates will help study the four-fatjet signal very easily.  For the sake of comparison, we go ahead and analyze our 
	four-fatjet signal for the previously considered benchmark points to show what sensitivity could be expected at the 
	LHC with a center-of-mass energy $\sqrt{s}=14$ TeV.
	
	Placing similar pre-selection criteria as in the previous cases, we choose a single set of cuts  to show our result:

	\noindent
	{\bf Cut-A :} 
\begin{enumerate}[i.]
		\item The $P_T$ of leading fatjet, ${P_T}_{j_1} > 500.0$ GeV and  sub-leading fatjet, ${P_T}_{j_2} > 300.0$ GeV. 
		\item Jet mass of leading jet, $m_{j_1} > 150.0$ GeV and sub-leading jet, $m_{j_2} > 120.0$ GeV while invariant mass of 
		leading and subleading jet, $m_{j_1 j_2} > 800.0$ GeV. Additionally, the invariant mass of all four fatjets, 
		$m_{j_1 j_2 j_3 j_4} > 1000.0$ GeV. 
		\item The $P_T$ of sub jets associated with leading charged lepton satisfies ${P_T}^{sj}_{l_1} > 200.0$ GeV.
\end{enumerate}
We summarize our cuts for four-fatjet inclusive searches in Tables~\ref{tab:cutflow4jet_summary}.

\begin{table*}[!hptb]
		\caption{ Cut summary for four-fatjet inclusive searches.}
		\label{tab:cutflow4jet_summary}
		\centering
		\renewcommand{\arraystretch}{1.5}
		\begin{tabular*}{\textwidth}{@{\extracolsep{\fill}}lllll@{}}\hline
			 &Cut summary & \\ \hline
		   {\bf Pre-selection cut} & {\bf Cut-A} \\
			\hline
			
  $\bullet$ \,\, $N_l \geq$ 2 with $P_T > 25$ GeV  
& $\bullet$ \,\, ${P_T}_{j_1} > 500$ GeV,${P_T}_{j_2} > 300$ GeV \\  
  $\bullet$ \,\, Two fatjets with radius $R = 0.8$  
& $\bullet$ \,\, $m_{j_i} > 150$ GeV ($i=1,2$), $m_{j_2} > 120$ GeV ($i=1,2$)\\
& $\bullet$ \,\, $m_{j_1 j_2} > 800$ GeV, $m_{j_1 j_2 j_3 j_4} > 1000$ GeV \\
  $\bullet$ \,\, $|\Delta \eta| < 4$ of two ungroomed fatjet 
& $\bullet$ \,\, ${P_T}^{sj}_{l_1} > 200$ GeV.\\
			\hline
		\end{tabular*}
	\end{table*}

		\begin{table}[!h]
		\caption{Signal significance of the four-fatjet signals  at the LHC with a center-of-mass energy $\sqrt{s}=14$ TeV with ${\cal L} = 3000$  fb$^{-1}$ and ${\cal L} = 600$  fb$^{-1}$.}
		\label{tablesignificance4jet}
		\centering
		\renewcommand{\arraystretch}{1.5}
		\begin{tabular*}{0.45\textwidth}{@{\extracolsep{\fill}}lll@{}}\hline
			After applying Cut & $\mathcal{L}=3000$  fb$^{-1}$  & $\mathcal{L}=600$  fb$^{-1}$\\
			\hline
			{\bf Cut-A} ({\tt BP1}) & $0.30$ $\sigma$ & $0.13$ $\sigma$\\
			\hline
			{\bf Cut-A} ({\tt BP2}) & $0.11$ $\sigma$ & $0.05$ $\sigma$\\
			\hline
			{\bf Cut-A} ({\tt BP3}) & $0.16$ $\sigma$ & $0.07$ $\sigma$\\
			\hline
		\end{tabular*}
		
	\end{table}
	
	As expected, we find that the four-fatjet analysis is not feasible with the available  cross section for the signal and, therefore, will not be accessible at the LHC, even with ${\cal L} = 3000$  fb$^{-1}$ as shown in Table~\ref{tablesignificance4jet}. This
	significance is further diminished if the $LSF$ and $LMD$ variables are invoked in our analysis.
	
	\section{Conclusion}\label{sec:conclusion}	
	In this work, we perform a collider study to look for fatjets originating from boosted particles that come from the decay of heavy resonances in a left-right symmetric extension of the SM. The model contains heavy leptons, heavy quarks, and heavy neutrinos, which play a fundamental role in achieving a universal seesaw mechanism for the generation of all the SM fermion masses. The fatjets, which include non-isolated leptons, originate from heavy neutrinos (few 100 GeV) that decay from a heavier (few TeV) right-handed gauge boson. We employ substructure-based variables $LSF$ and $LMD$, together with hard kinematic cuts, to search for fatjets with associated charged leptons, to reduce SM background while keeping enough statistics for the signals.
	
	We analyze a multi-fatjet signal topology that would be typical in the model when the heavy fermionic states are produced and finally decay to SM particles. We have focused mainly on two-fatjet and three-fatjet final state searches while commenting on the possibility of a four-fatjet scenario with very limited signal sensitivity for our parameter choices. A critical aspect of the analysis included the presence of a charged lepton in the fatjets, helping us identify the signal over the SM background by employing the variables $LSF$ and $LMD$.
	
	In the two-fatjet final state, both jets include a charged lepton, while for three-fatjet, two of them include a charged lepton. We find that the two-fatjet signals, which also include leptons, can be discovered at the LHC with a moderate luminosity of ${\cal L}\simeq 600$ fb$^{-1}$ for $m_E\simeq 1$ TeV, $m_{W_R}= 4.5$ TeV ({\tt BP1}) and $m_E\simeq 0.72$ TeV, $m_{W_R}= 5$ TeV ({\tt BP3}), while the signal can be discovered with ${\cal L}\simeq 3000$ fb$^{-1}$ for $m_E\simeq 1$ TeV, $m_{W_R}\simeq 5$ TeV ({\tt BP2}), as shown in Tables~\ref{tab:significance2jet}. On the other hand, a three-fatjet signal can be discovered with ${\cal L}\simeq 600$ fb$^{-1}$ (see Table~\ref{significance3jet}) for $m_{W_R}= 4.5$ TeV ({\tt BP1}). In the case of $m_{W_R}= 5$ TeV ({\tt BP2} and {\tt BP3}), we require ${\cal L}\simeq 3000$ fb$^{-1}$ for a $5\sigma$ discovery.

 We finally conclude by noting that although these signals of fatjet can have origins from an altogether different underlying theoretical framework compared to left-right symmetry or a seesaw framework, their observation will be still crucial in the search of new physics at LHC and will 
 provide hints on the presence of new heavy states beyond the SM.

\section*{ACKNOWLEDGEMENT}
	The authors would like to acknowledge support from the Department of Atomic Energy, Government of India, for the Regional Centre for Accelerator-based Particle Physics (RECAPP), Harish Chandra Research Institute.

\appendix
 \section{Decay branching ratios in {\tt BP2} and 
 {\tt BP3}}\label{app:branching}
 We present the relevant branching ratios (BR) for the decay modes of the heavy $SU(2)_R$ gauge bosons $W_R$ and $Z_R$, the heavy leptons $E_i$, and three heavy neutrinos $\nu_i$ (with $i=4,5,6$) for {\tt BP2} and {\tt BP3} in Table~\ref{tab:branchings-BP23}. The BRs of the particles in {\tt BP1} and {\tt BP2} remain roughly the same due to the identical mass of the heavy charged leptons. In contrast, there are subtle changes on the BRs in {\tt BP3} due to the smaller mass of the heavy charged leptons. Among the relevant BRs, the di-jet BR of the heavy gauge bosons are reduced by approximately 2\% to 3\%, ${\rm BR}(W_R\to e\nu_4)$ is reduced by roughly 2\%, while ${\rm BR}(W_R\to E_4\nu_4)$ is increased by 2\%. The other BRs in {\tt BP3} are roughly the same as those in {\tt BP1}.
 	\begin{table*}[ht!]
		\caption{\label{tab:branchings-BP23} Decay branching ratios for $W_R$ and $Z_R$, heavy charged leptons $E_i$, and the lightest three heavy neutrinos $\nu_i$, ($i=4,5,6$) for {\tt BP2} and {\tt BP3}. }
		\centering	
		\renewcommand{\arraystretch}{1.3}
		\begin{tabular*}{0.3\textwidth}{@{\extracolsep{\fill}}|lll@{}}\hline
			Decay & \multicolumn{2}{c}{Branching} \\ \hline
                & {\tt BP2} & {\tt BP3} \\ \hline
			$W_R^+\to jj$ & 65.0\% & 63.0\% \\
			$W_R^+\to t\bar{b}$ & 10.0\% & 9.80\%\\
			$W_R^+\to e^+\nu_4$ &   4.96\% &3.20\% \\
			$W_R^+\to \mu^+\nu_5$ &   1.56\% &0.700\% \\
			$W_R^+\to \tau^+\nu_6$ &   2.40\% & 1.40\% \\
			$W_R^+\to E_4^+\nu_4$ &   5.20\% & 7.00\% \\
			$W_R^+\to E_5^+\nu_6$ &   5.48\% &7.21\%\\
			$W_R^+\to E_6^+\nu_5$ &   5.30\% & 7.31\%\\\hline
		\end{tabular*}	
		\begin{tabular*}{0.3\textwidth}{@{\extracolsep{\fill}}|lll@{}}\hline
			Decay & \multicolumn{2}{c}{Branching} \\ \hline
                & {\tt BP2} & {\tt BP3} \\ \hline
                $Z_R\to jj$ & 46.9\%  & 45.9\%\\  
			$Z_R\to \nu_4\nu_4$ & 7.09\% & 6.94\% \\ 
			$Z_R\to \nu_5\nu_5$ & 6.52\% & 6.38\% \\
			$Z_R\to E_4\bar{E_4}$ & 2.27\% & 2.40\% \\  
			$Z_R\to E_5\bar{E_5}$ & 1.41\% & 2.28\% \\  
			&& \\
			&& \\
			&& \\ \hline
		\end{tabular*}	
		\begin{tabular*}{0.3\textwidth}{@{\extracolsep{\fill}}|lll|@{}}\hline
			Decay & \multicolumn{2}{c|}{Branching} \\ \hline
                & {\tt BP2} & {\tt BP3} \\ \hline
			$E_4\to \nu_4 jj$ & 70.7\%& 70.4\% \\
			$E_5\to \nu_6 jj$ & 5.88\%& 0.712\% \\
			$E_6\to \nu_5 jj$ & 73.7\%& 77.9\% \\ 
			$E_6\to \mu jj$ & 3.66\%& 2.18\% \\ 			
			$\nu_4\to e^\pm jj$ & 100\%& 100\% \\
			$\nu_5\to \mu^\pm jj$ & 94.5\%& 95.0\% \\
			$\nu_6\to \tau^\pm jj$ & 87.4\%& 88.6\% \\ 
			&& \\\hline
		\end{tabular*}	
		
	\end{table*}		

\section{Kinematical distributions for  fatjets in three-fatjet searches}\label{app:dist}
\begin{figure*}[!h]
		\centering
		\includegraphics[width=0.495\textwidth]{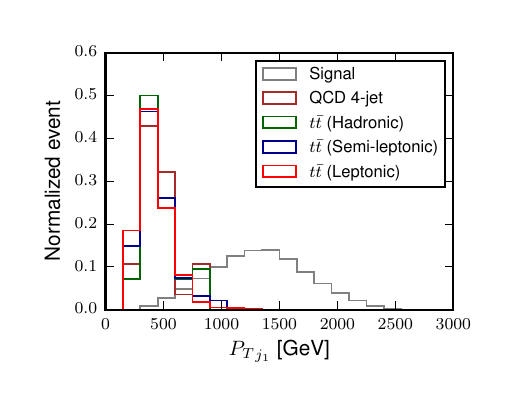}
		\includegraphics[width=0.495\textwidth]{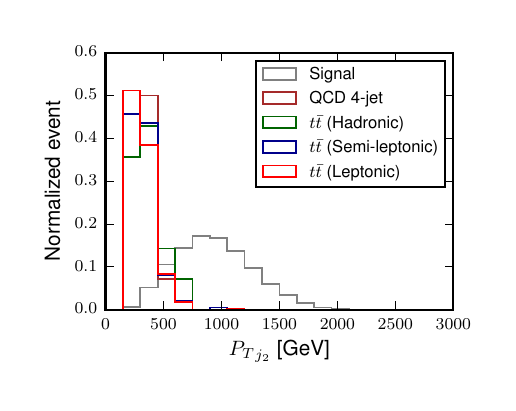} \\
		\includegraphics[width=0.495\textwidth]{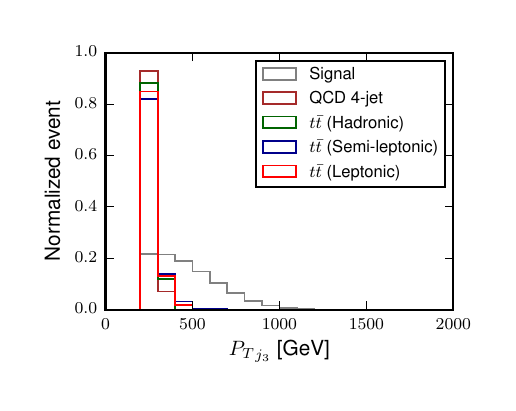}
		\includegraphics[width=0.495\textwidth]{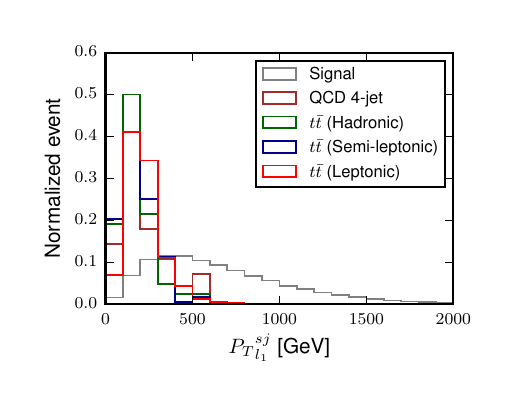} \\
		\includegraphics[width=0.495\textwidth]{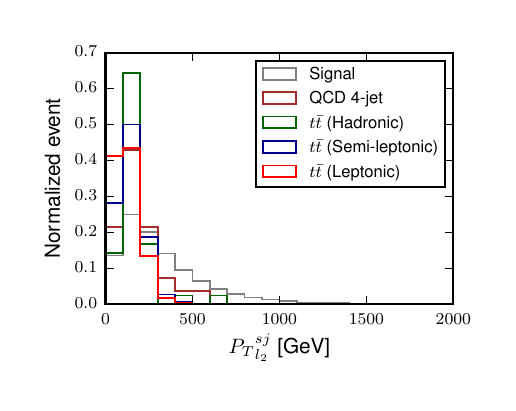} 
		\includegraphics[width=0.495\textwidth]{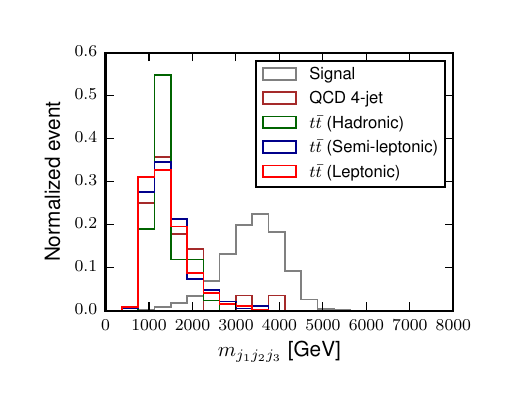}
		\caption{Normalized distributions of the signal and different SM backgrounds as a function of the transverse momentum ($P_T$) of the three leading fatjets,  the transverse momentum of the leptonic sub-jets ($P_{T_{l}^{sj}}$) in the fatjets and invariant mass ($m_{j_1 \, j_2 \ j_3}$) of the three leading fatjets in the {\it three-fatjet inclusive} searches for {\tt BP2}.}
		\label{dist_3jet}
	\end{figure*}

\begin{figure*}[!h]
		\centering
		\includegraphics[width=0.495\textwidth]{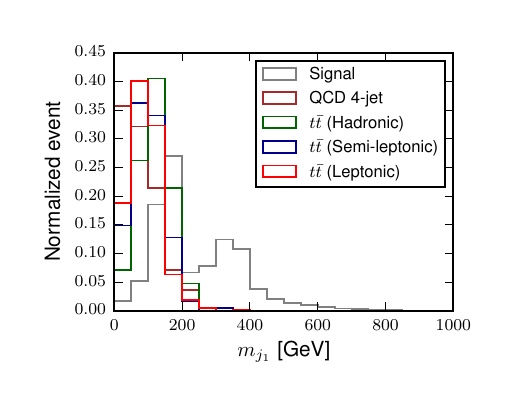}
		\includegraphics[width=0.495\textwidth]{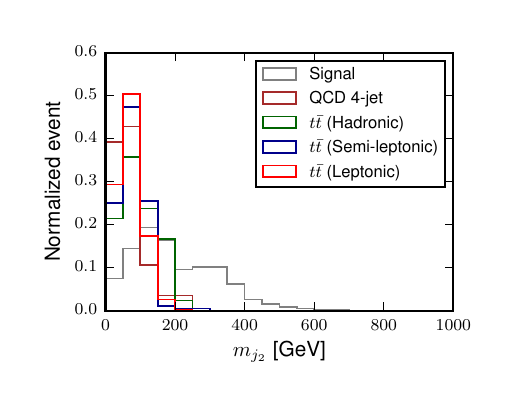}\\ 
		\includegraphics[width=0.495\textwidth]{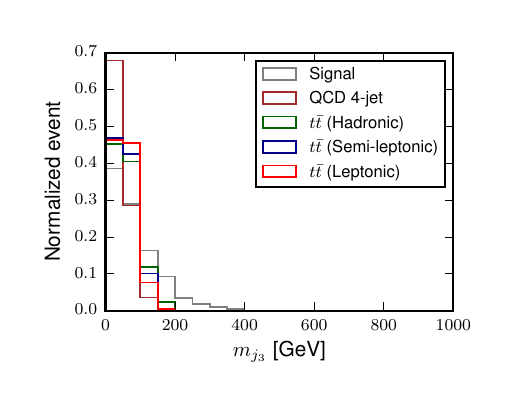}
		\includegraphics[width=0.495\textwidth]{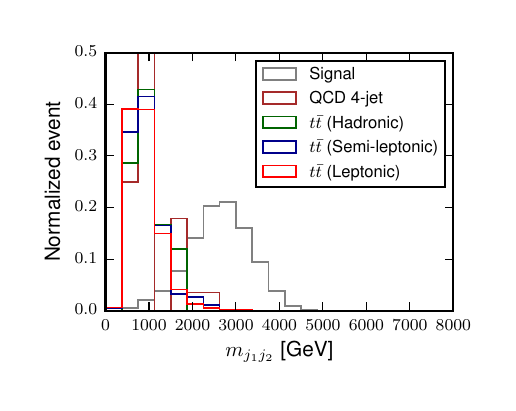} \\
		\includegraphics[width=0.495\textwidth]{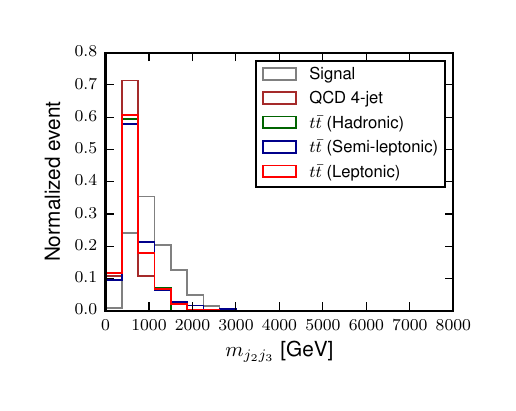} 
		\includegraphics[width=0.495\textwidth]{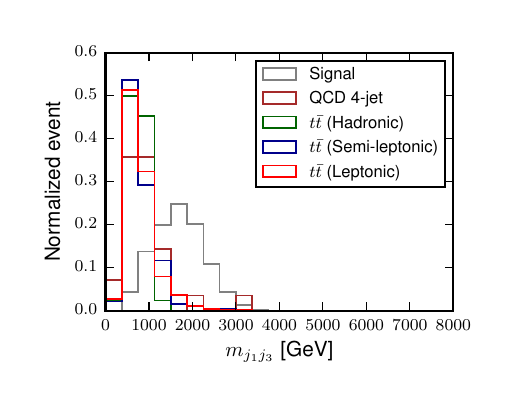}
			\caption{Normalized distributions of the signal and different SM backgrounds as a function of the 
   mass ($m_{j_i}$) of three leading fatjets and the invariant mass ($m_{j_a \, j_b}$, a,b=1,2,3) in the {\it three-fatjet inclusive} searches 
   for {\tt BP2}.}
		\label{dist1_3jet}
	\end{figure*}

\begin{figure*}[!h]
		\centering
		\includegraphics[width=0.495\textwidth]{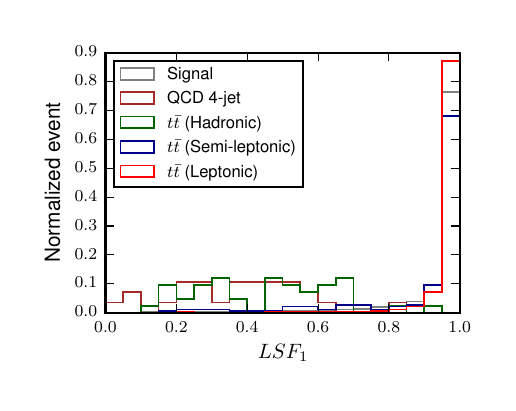}
		\includegraphics[width=0.495\textwidth]{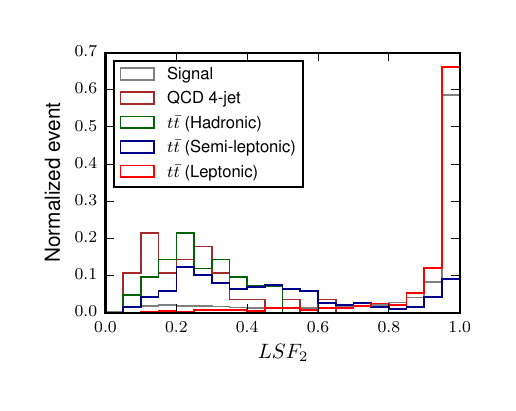}\\ 
		\includegraphics[width=0.495\textwidth]{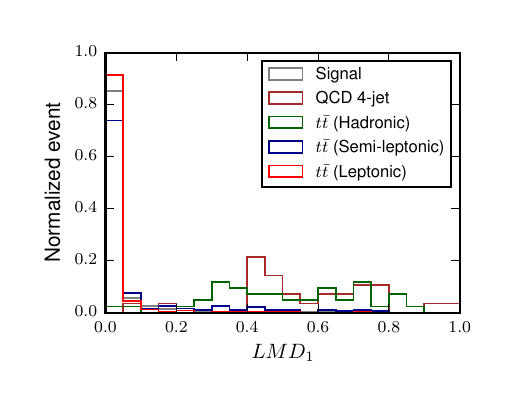}
		\includegraphics[width=0.495\textwidth]{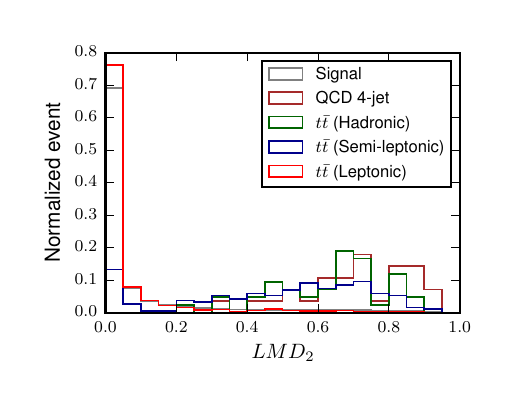} 
			\caption{Normalized distributions of the signal and different SM backgrounds as a function of $LSF$ and $LMD$ in the {\it three-fatjet inclusive} searches 
   for {\tt BP2}.}
		\label{dist2_3jet}
	\end{figure*}

The normalized distributions for the kinematic variables, such as the $P_T$ of the fatjets, $P_T$ of the two leading lepton subjects, invariant mass of the fatjets, the $LSF$ and $LMD$ in the case of three-fatjet search are shown in Figs.~\ref{dist_3jet}, \ref{dist1_3jet} and \ref{dist2_3jet}. The distributions are similar to the distributions 
 of two-fatjet searches as described above in the main texts. In particular, the invariant mass constructed out of the three fatjets shown in Fig.~\ref{dist_3jet}, is identified as a good observable to separate 
 signals from backgrounds. 
 \section{Signal and background events surviving after different cuts for different benchmark points}\label{app:tables}
 
We list the number of signal events and background events after applying the pre-selection cuts and different selection cuts in Tables~\ref{tab:cutflow2jet_bp2} and \ref{tab:cutflow2jet_bp3} for two-fatjet searches in {\tt BP2} and {\tt BP3}, respectively. For three-fatjet searches, the signal and background events after different cuts are listed in Tables~\ref{tablecutflow3jet_bp1}, \ref{tablecutflow3jet_bp2}, and \ref{tablecutflow3jet_bp3} for {\tt BP1}, {\tt BP2},  and {\tt BP3}, respectively.
 
 	\begin{table*}[!hptb]
		\caption{Signal and background events surviving after applying the pre-selection criteria as well as selection cuts at $\sqrt s = 14$ TeV and ${\cal L} = 3000$  fb$^{-1}$ for two-fatjet searches in {\tt BP2}. The individual contributions of SM subprocesses to the background are as shown in Table~\ref{tab:cutflow2jet_bp1}.} 
		\label{tab:cutflow2jet_bp2}
		\centering
		\renewcommand{\arraystretch}{1.5}
		\begin{tabular*}{\textwidth}{@{\extracolsep{\fill}}lllll@{}}\hline
			Data sets&Cross section (pb)&\multicolumn{3}{c}{Number of events}\\ \hline
			& Parton-level & Pre-selection cut & {\bf Cut-A} & {\bf Cut-B} \\
			\hline
			$W_R \to E_6 \nu_5$, $E_6 \to \mu j j$  & $0.009424 \times 10^{-3}$ & $4$  & $1$ & $1$   \\
			\hline
			$W_R \to E_6 \nu_5$, $E_6 \to \nu_5 j j$  & $0.18038 \times 10^{-3}$ & $146$  & $46$ & $83$   \\
			\hline
			$W_R \to E_4 \nu_4$ & $0.19974 \times 10^{-3}$ & $286$  & $22$ & $93$   \\
			\hline
			$Z_R \to (\nu_4 \nu_4)/(\nu_5 \nu_5)$  & $0.04861 \times 10^{-3}$ & $75$  & $15$ & $34$   \\
			\hline
			$Z_R \to E_4 \bar{E_4}$ & $0.004374 \times 10^{-3}$ & $4$  & $0$ & $1$ \\
			\hline
			Total Signal ($s$)&& $515$ & $84$ & $212$ \\ \hline
			Total Background ($b$) & & $1.7\times 10^7 $& $105 $& $1.568\times 10^3 $ \\ \hline				
		\end{tabular*}
				
	\end{table*}
	%
	\begin{table*}[!hptb]
		\caption{Signal and background events surviving after applying the pre-selection criteria as well as selection cuts at $\sqrt s = 14$ TeV and ${\cal L} = 3000$  fb$^{-1}$ for two-fatjet searches in {\tt BP3}. The individual contributions of SM subprocesses to the background are as shown in Table~\ref{tab:cutflow2jet_bp1}.} 
		\label{tab:cutflow2jet_bp3}
		\centering
		\renewcommand{\arraystretch}{1.5}
		\begin{tabular*}{\textwidth}{@{\extracolsep{\fill}}lllll@{}}\hline
			Data sets&Cross section (pb)&\multicolumn{3}{c}{Number of events}\\ \hline
			& Parton-level & Pre-selection cut & {\bf Cut-A} & {\bf Cut-B} \\
			\hline
			$W_R \to e_6 \nu_5$, $e_6 \to \mu j j$  & $0.009\times 10^{-3}$ & 2 & 1 & 1   \\
			\hline
			$W_R \to e_6 \nu_5$, $e_6 \to \nu_5 j j$  & $0.279 \times 10^{-3}$ & 271 & 83 & 152   \\
			\hline
			$W_R \to E_4 \nu_4$ & $0.295 \times 10^{-3}$ & 519 & 34 & 125   \\
			\hline
			$Z_R \to (\nu_4 \nu_4)/(\nu_5 \nu_5)$  & $0.048 \times 10^{-3}$ & 73 & 14 & 33 \\
			\hline
			$Z_R \to E_4 \bar{E_4}$ & 0.006 fb & 7 & 1 & 1 \\
			\hline
			Total Signal ($s$)&& $872$ & $133$ & $312$ \\ \hline
			Total Background ($b$) & & $1.7\times 10^7 $& $105 $& $1.568\times 10^3 $ \\ \hline				
		\end{tabular*}
	\end{table*}
	%
	\begin{table*}[!hptb]
		\caption{Signal and background events surviving after applying the pre-selection criteria as well as selection cuts at $\sqrt s = 14$ TeV and ${\cal L} = 3000$  fb$^{-1}$ for three-fatjet searches in {\tt BP1}.} 
		\label{tablecutflow3jet_bp1}		
		\centering
		\renewcommand{\arraystretch}{1.5}
		\begin{tabular*}{\textwidth}{@{\extracolsep{\fill}}llllll@{}}\hline
			Data sets&Cross section (pb)&\multicolumn{4}{c}{Number of events}\\ \hline
			& Parton-level & Pre-selection cut & {\bf Cut-A} & {\bf Cut-B} & {\bf Cut-C}\\ \hline
			$W_R \to e_6 \nu_5$ & $0.400\times 10^{-3}$ & 378 & 51 & 88 & 49   \\
			\hline
			$W_R \to E_4 \nu_4$ & $0.485 \times 10^{-3}$ & 607 & 22 & 104 & 21  \\
			\hline
			$Z_R \to E_4 \bar{E_4}$ & $ 0.012 \times 10^{-3}$ & 11 & 1 & 1 & 1 \\
			\hline
			Total Signal ($s$) && $996$ & $74$ & $193$ & $ 114$\\ \hline				
			QCD $4$-jet  & $90387.504$ & $7.59\times 10^6$ & $0$& $0$ & $0$\\
			\hline
			$t \bar{t}$ (Hadronic) & $229.603$ & $2.89\times 10^4$  & $0$ &  $0$ & $0$\\
			\hline
			$t \bar{t}$ (Semi-leptonic) & $178.569$ & $1.01\times 10^5$  & $536$ & $0$  & $536$  \\
			\hline
			$t \bar{t}$ (Leptonic) &  $34.854$ & $7.9\times 10^4$ & $0$ &  $105$ & $0$\\
			\hline
			Total Background ($b$) & & $7.8\times 10^6 $& $536 $& $105$ & $536 $ \\ \hline				
		\end{tabular*}
	\end{table*}
	\begin{table*}[!hptb]
		\caption{Signal and background events surviving after applying the pre-selection criteria as well as selection cuts at $\sqrt s = 14$ TeV and ${\cal L} = 3000$  fb$^{-1}$ for three-fatjet searches in {\tt BP2}. The individual contributions of SM subprocesses to the background are as shown in Table~\ref{tablecutflow3jet_bp1}.} 
		\label{tablecutflow3jet_bp2}
		\renewcommand{\arraystretch}{1.5}
		\begin{tabular*}{\textwidth}{@{\extracolsep{\fill}}llllll@{}}\hline
			Data sets&Cross section (pb)&\multicolumn{4}{c}{Number of events}\\ \hline
			& Parton-level & Pre-selection cut & {\bf Cut-A} & {\bf Cut-B} & {\bf Cut-C}\\ \hline
			$W_R \to E_4 \nu_4$  & $0.199 \times 10^{-3}$ & $251$ &  $9$ & $42$ & $8$\\
			\hline
			$W_R \to E_6 \nu_5$  & $0.180 \times 10^{-3}$ & $207$ & $28$ & $49$ & $28$ \\
			\hline
			$Z_R \to E_4 \bar{E_4}$  & $0.004 \times 10^{-3}$ & $4$ &  $0$ & $0$  & $0$\\  \hline
			Total Signal ($s$) && $462$ & $37$ & $91$ & $ 36$\\ \hline					
			Total Background ($b$) & & $7.8\times 10^6 $& $536 $& $105$& $536 $ \\ \hline	
		\end{tabular*}
	\end{table*}
	\begin{table*}[!hptb]
		\caption{Signal and background events surviving after applying the pre-selection criteria as well as selection cuts at $\sqrt s = 14$ TeV and ${\cal L} = 3000$  fb$^{-1}$ for three-fatjet searches in {\tt BP3}. The individual contributions of SM subprocesses to the background are as shown in Table~\ref{tablecutflow3jet_bp1}.} 
		\label{tablecutflow3jet_bp3}
		\centering
		\renewcommand{\arraystretch}{1.5}
		\begin{tabular*}{\textwidth}{@{\extracolsep{\fill}}llllll@{}}\hline
			Data sets&Cross section (pb)&\multicolumn{4}{c}{Number of events}\\ \hline
			& Parton-level & Pre-selection cut & {\bf Cut-A} & {\bf Cut-B} & {\bf Cut-C}\\ \hline
			$W_R \to e_6 \nu_5$ & $0.279\times 10^{-3}$ & 313 & 46 & 79 & 45  \\
			\hline
			$W_R \to E_4 \nu_4$ & $0.295 \times 10^{-3}$ & 162 & 3 & 17 & 3  \\
			\hline
			$Z_R \to E_4 \bar{E_4}$ &$ 0.006 \times 10^{-3}$ & 2 & 0 & 0 & 0 \\
			\hline
			Total Signal ($s$) && $477$ & $49$ & $96$ & $ 48$\\ \hline					
			Total Background ($b$) & & $7.8\times 10^6 $& $536 $& $105$& $536 $ \\ \hline	
		\end{tabular*}
		\end{table*}
\clearpage
	\bibliographystyle{utphysM}
	\bibliography{References}
\end{document}